\documentclass[preprint,prd,aps,showpacs,showkeys,nofootinbib]{revtex4}
\usepackage{graphicx}
\begin{document}

\title{\boldmath ${B^0} - {{\bar B}^0}$ mixing  in supersymmetry with gauged baryon and lepton numbers }

\author{Fei Sun$^{a,b}$\footnote{Corresponding author. email:sunfei@mail.dlut.edu.cn}, Tai-Fu Feng$^{a,b,c,d}$, Shu-Min Zhao$^{a,c,d}$, Hai-Bin Zhang$^{a,b}$, Tie-Jun Gao$^{c}$, Jian-Bin Chen$^{e}$}

\affiliation{$^a$Department of Physics, Hebei University, Baoding, 071002, China \\
$^b$Department of Physics, Dalian University of Technology, Dalian, 116024, China \\
$^c$Institute of theoretical Physics, Chinese Academy of Sciences,  Beijing, 100190, China \\
$^d$The Key Laboratory of Mathematics-Mechanization (KLMM), Beijing, 100190, China \\
$^e$Department of Physics, Taiyuan University of Technology, Taiyuan, 030024, China}

\begin{abstract}

We perform an analysis on ${B^0} - {{\bar B}^0}$ mixing in the extension of the minimal supersymmetric standard model where baryon and lepton numbers are local gauge symmetries (BLMSSM) by using the effective Hamiltonian method. And the constraint of a 125 GeV Higgs to the parameter space has also been considered. The numerical results indicate that the contributions of the extra particles can be sizeable in ${B^0} - {{\bar B}^0}$ mixing.
For certain parameter sets,  the theoretical prediction of  mass differences $\Delta m_{B}$
agrees with the current experimental result. Furthermore,  ${B^0} - {{\bar B}^0}$ mixing in the BLMSSM can preliminarily constrain the parameter space. With  the development of more precise theoretical analysis and experimental determinations,  the  ${B^0} - {{\bar B}^0}$ mixing in the BLMSSM  will   have a  clearer picture and  the parameter space in this model will also be further constrained.
\end{abstract}



\maketitle

\section{Introduction\label{sec1}}
The Minimal Supersymmetric Standard Model (MSSM)~\cite{H.P. Nilles,H.E. Haber and G.L. Kane,J. Rosiek,H.E. Haber,S.P. Martin}, as one of the most appealing options for the   physics beyond the Standard Model (SM), has  drawn the physicists' attention  for a long time. As the simplest soft broken supersymmetry (SUSY) theory,  the MSSM can solve hierarchy problem, ensure that the gauge couplings unify at high energies and provide a good dark matter candidate.
To search for new particles predicted by SUSY, the Large Hadron Collider (LHC) has collected huge amounts of data,
the CMS~\cite{CMS0} and ATLAS~\cite{ATLAS0} experiments now set strong limits on these parameter space~\cite{P. Bechtle1,B. C. Allanach,O. Buchmuller,P. Bechtle2}. However, the present searches are largely based on the assumption of conserved R-parity~\cite{G. R. Farrar and P. Fayet}.
Some studies in the  low-energy SUSY have been motivated by the results of the LHC~\cite{G. Kane,H. Baer,J. L. Feng,S. Heinemeyer,A. Arbey,A. Arbey1,P. Draper,U. Ellwanger,S. Akula,M. Kadastik,Junjie Cao},
and R-parity violating scenarios of general MSSM have been proposed~\cite{C.S. Aulakh,L.J. Hall,I.H. Lee,S. Dawson,G.G. Ross,R. Barbieri,S. Dimopoulos,R. Hempfling,J. Erler,S. Roy,H.P. Nilles1,H. K. Dreiner,M.A. D,M. Hirsch,F. de Campos,C.-H. Chang,C. Csaki,B. C. Allanach and B. Gripaios,H. K. Dreiner1,P. W. Graham,M. Hanussek and J. S. Kim,H. Dreiner,ref-zhang1,ref-zhang2,ref-zhang3}.

A model based on the gauge symmetry group $SU(3) \otimes SU(2) \otimes U{(1)_Y} \otimes U{(1)_B} \otimes U{(1)_L}$ has been investigated  at the TeV scale recently \cite{P. Fileviez,T. R. Dulaney,P. Fileviez Perez,P. Ko},
where $B$ stands for baryon number and $L$ stands for  lepton number. In this theory, the baryon and lepton numbers are local gauge  symmetries spontaneously broken at the TeV scale.
Breaking baryon number can explain the origin of the matter-antimatter asymmetry in the Universe.  And breaking lepton number can explain the smallness of neutrino masses \cite{P.Minkowski,T.Yanagida,M.Gell-Mann,S.L.Glashow,R.N.Mohapatra}.
Two extensions of the  SM  where  $B$ and $L$ are spontaneously broken gauge symmetries near the weak scale are constructed \cite{P. F. Perez}:  model I
is a non-supersymmetric extension~\cite{R.Foot,C.D.Carone}; model II (BLMSSM) is a supersymmetric extension and is more  favoured by  the experiments \cite{P. F. Perez. Phys. Lett}. The BLMSSM has been studied in great detail and could avoid the current LHC bounds on the SUSY mass spectrum \cite{P. F. Perez and M. B. Wise1,P. F. Perez and M. B. Wise2,J. M. Arnold}. Some further  phenomenology analysis based on the BLMSSM  coincide with the current experimental data well, the mass and decays of the lightest CP-even Higgs have been investigated in Refs. \cite{J. M. Arnold,Tai-Fu Feng}, and the neutron electric dipole moment in CP violating BLMSSM has also been studied  \cite{Shu-Min Zhao}.

The flavor changing neutral current (FCNC) processes are highly suppressed in the SM, therefore it is a fertile ground to search for  physics beyond SM (BSM).
FCNC processes such as  $b \to s\gamma$, ${K^0} - {{\bar
K}^0}$ and ${B^0} - {{\bar B}^0}$ mixing have played an important
role in particle physics over the last four decades. It is well known that CP violation was first observed in the
decays of $K_L^0$ meson in 1964 \cite{cp}, and CP violation of the
neutral $B$ meson  system was observed in 2001 \cite{K. Abe}.
The first indication of a large top quark mass was also given by ${B^0} - {{\bar B}^0}$ mixing ~\cite{C. Albajar,J. R. Ellis0}.  $B$-system decays have an advantage over the
${K}$-system to provide a direct test of the CP violating of SM
and is free of corrections from strong interactions \cite{K. Abe et al,B. Aubert et al0,T. Hurth et al}.
The experiment results of ${B^0} - {{\bar B}^0}$ mixing have been published
by the ALEPH \cite{D. Buskulic et al}, DELPHI \cite{J. Abdallah et al,P. Abreu et al}, L3 \cite{M. Acciarri et al}, OPAL \cite{G. Alexander et al,K. Ackerstaff et al}
BaBar \cite{B. Aubert et al}, Belle \cite{N.C. Hastings et al}, CDF \cite{F. Abe et al}, D{\O} \cite{V.M. Abazov et al}, and LHCb \cite{R. Aaij et al}
collaborations. Current
experimental result of mass difference is $\Delta m_B^{Exp} = 0.507 \pm 0.004 ~\rm{ps^{ - 1}}= \left({3.337 \pm 0.033}
\right) \times {10^{ - 13}}~{\rm{GeV}}$ \cite{PDG}. Calculations for
${B^0} - {{\bar B}^0}$ mixing have been done in
the SM , the two-Higgs doublet model (2HDM), the MSSM and other models \cite{smburas,John S. HAGELIN,J. Urban,fengtaifu,F. Krauss,Soo-hyeon Nam,Debrupa Chakraverty,Javier Virto,FEI SUN,FEI SUN1}. The SM prediction for mass difference is
$\Delta m_B^{SM} = 0.543 \pm 0.091 ~\rm{ps^{ - 1}}$ \cite{A. Lenz and U. Nierste}, which has a good agreement with the experiment.  However,  the theoretical error  is around
17\%,  which is considerably larger than the experimental error.
The running of LHC will resume in 2015 with higher energy and luminosity.
Proposals for next-generation B-factories including SuperKEKB  in Japan whose target  luminosity is $8 \times
{10^{35}}~\rm{c{m^{ - 2}}{s^{ - 1}}}$ will start collecting data in the near
future \cite{kekb}. This may also give some hints on physics beyond
the SM. So it  is  important  for experimental and theoretical
physicist to search for new physics. As a candidate of new physics, the BLMSSM
provides new FCNC at loop level in the  ${B^0} - {{\bar B}^0}$
mixing. We will carry out
our calculations for ${B^0} - {{\bar B}^0}$ mixing in this model.

Our presentation is organized as follows. In Section~\ref{sec2}, we briefly summarize the main features of the BLMSSM and introduce the superpotential as well as  soft breaking terms,   then  we obtain the mass matrices and couplings needed for ${B^0} - {{\bar B}^0}$ mixing.
In Section~\ref{sec3}, we give the analytical formulae of the ${B^0} - {{\bar B}^0}$ mixing in BLMSSM. The numerical analysis are shown in Section~\ref{sec4}.  Section~\ref{sec5} presents our conclusions.  Finally, some related formulae are given in Appendix~\ref{Integral}--\ref{hadronic}.

\section{BLMSSM\label{sec2}}
In this section, we briefly review some  main features of the BLMSSM. In the  BLMSSM with gauged baryon $(B)$ and lepton $(L)$, by adding the new quarks with baryon number $B_{4}=\frac{3}{2}$ and
the new leptons with lepton number $L_{4}=\frac{3}{2}$, one can cancel the  baryonic and leptonic anomalies respectively \cite{P. F. Perez}.
Compared with the MSSM, the BLMSSM  includes many new fields. Tables \ref{quarks}--\ref{stability} list the superfields including the new quarks, new leptons, new Higgs, the exotic superfields $\hat{X}$ and $\hat{X'}$, respectively. As one can see, the left-handed superfields have the same absolute value of $U(1)_B$ as that of  the right-handed superfields  but with a contrary sign  to cancel baryonic  anomalies in the quark sector, similarly for the  $U(1)_L$  in the leptonic sector  to cancel leptonic anomalies.

\begin{table}[htdp]
\caption{Superfields including the new quarks in the BLMSSM.}
\begin{center}
\begin{tabular}{|c|c|c|c|c|c|}
\hline
Superfields & $SU(3)_C$ & $SU(2)_L$ & $U(1)_Y$ & $U(1)_B$ & $U(1)_L$\\
\hline
\hline
$\hat{Q}_4$ & 3 & 2 & 1/6 & $B_4$ & 0 \\
\hline
$\hat{U}^c_4$ & $\bar{3}$ & 1 & -2/3 & -$B_4$ & 0 \\
\hline
$\hat{D}^c_4$ & $\bar{3}$ & 1 & 1/3 & -$B_4$ & 0 \\
\hline
$\hat{Q}_5^c$ & $\bar{3}$ & 2 & -1/6 & -$(1+B_4)$ & 0 \\
\hline
$\hat{U}_5$ & $3$ & 1 & 2/3 &  $1 + B_4$ & 0 \\
\hline
$\hat{D}_5$ & $3$ & 1 & -1/3 & $1 + B_4$ & 0 \\
\hline
\end{tabular}
\end{center}
\label{quarks}
\end{table}%

\begin{table}[htdp]
\caption{Superfields including the new leptons in the BLMSSM.}
\begin{center}
\begin{tabular}{|c|c|c|c|c|c|}
\hline
Superfields & $SU(3)_C$ & $SU(2)_L$ & $U(1)_Y$ & $U(1)_B$ & $U(1)_L$\\
\hline
\hline
$\hat{L}_4$ & 1 & 2 & -1/2 & 0 & $L_4$ \\
\hline
$\hat{E}^c_4$ & 1 & 1 & 1 & 0 & -$L_4$ \\
\hline
$\hat{N}^c_4$ & 1 & 1 & 0 & 0 & -$L_4$ \\
\hline
$\hat{L}_5^c$ & 1 & 2 & 1/2 & 0 & -$(3 + L_4)$ \\
\hline
$\hat{E}_5$ & 1 & 1 & -1 & 0 & $3 + L_4$ \\
\hline
$\hat{N}_5$ & 1 & 1 & 0 & 0 & $3 + L_4$ \\
\hline
\end{tabular}
\end{center}
\label{leptons}
\end{table}%

\begin{table}[htdp]
\caption{Superfields including the new Higgs in the BLMSSM.}
\begin{center}
\begin{tabular}{|c|c|c|c|c|c|}
\hline
Superfields & $SU(3)_C$ & $SU(2)_L$ & $U(1)_Y$ & $U(1)_B$ & $U(1)_L$\\
\hline
\hline
$\hat{\Phi}_B$ & 1 & 1 & 0 & 1 & 0 \\
\hline
$\hat{\varphi}_B$ & 1 & 1 & 0 & -1 & 0 \\
\hline
$\hat{\Phi}_L$ & 1 & 1 & 0 & 0 & -2 \\
\hline
$\hat{\varphi}_L$ & 1 & 1 & 0 & 0 & 2 \\
\hline
\end{tabular}
\end{center}
\label{Higgs}
\end{table}%

\begin{table}[htdp]
\caption{Superfields avoiding stability for the exotic quarks in the BLMSSM.}
\begin{center}
\begin{tabular}{|c|c|c|c|c|c|}
\hline
Superfields & $SU(3)_C$ & $SU(2)_L$ & $U(1)_Y$ & $U(1)_B$ & $U(1)_L$\\
\hline
\hline
$\hat{X}$ & 1 & 1 & 0 & $2/3 + B_4$ & 0 \\
\hline
$\hat{X'}$ & 1 & 1 & 0 & $-(2/3 + B_4)$ & 0 \\
\hline
\end{tabular}
\end{center}
\label{stability}
\end{table}%


In order to break baryon number spontaneously,  we need to introduce the superfields    $\hat{\Phi}_B$ and $\hat{\varphi}_B$ to acquire nonzero vacuum expectation values (VEVs),   which also generate  large mass for the new quarks. Similarly, we introduce the superfields $\hat{\Phi}_L$ and $\hat{\varphi}_L$ to acquire VEVs spontaneously breaking lepton number. Finally, the exotic quarks should be  unstable, so the model also includes the superfields $\hat{X}$ and $\hat{X'}$ to avoid the stability for the exotic quarks. Here $\hat{\Phi}_B$ and $\hat{\varphi}_B$ have $U(1)_B$ charge 1 and -1, respectively,  $\hat{\Phi}_L$ and $\hat{\varphi}_L$ have $U(1)_L$ charge -2 and 2, respectively. For superfields $\hat{X}$ and $\hat{X'}$, $U(1)_B$ charge is $2/3 + B_4$ and $-(2/3 + B_4)$, respectively. Here the lightest $X$ could be a dark matter candidate.

The superpotential in BLMSSM is written as
\begin{eqnarray}
&&{\cal W}_{_{BLMSSM}}={\cal W}_{_{MSSM}}+{\cal W}_{_B}+{\cal W}_{_L}+{\cal W}_{_X}\;,
\label{superpotential1}
\end{eqnarray}
where ${\cal W}_{_{MSSM}}$ is the superpotential of  MSSM, and
\begin{eqnarray}
&&{\cal W}_{_B}=\lambda_{_Q}\hat{Q}_{_4}\hat{Q}_{_5}^c\hat{\Phi}_{_B}+\lambda_{_U}\hat{U}_{_4}^c\hat{U}_{_5}
\hat{\varphi}_{_B}+\lambda_{_D}\hat{D}_{_4}^c\hat{D}_{_5}\hat{\varphi}_{_B}+\mu_{_B}\hat{\Phi}_{_B}\hat{\varphi}_{_B}
\nonumber\\
&&\hspace{1.2cm}
+Y_{_{u_4}}\hat{Q}_{_4}\hat{H}_{_u}\hat{U}_{_4}^c+Y_{_{d_4}}\hat{Q}_{_4}\hat{H}_{_d}\hat{D}_{_4}^c
+Y_{_{u_5}}\hat{Q}_{_5}^c\hat{H}_{_d}\hat{U}_{_5}+Y_{_{d_5}}\hat{Q}_{_5}^c\hat{H}_{_u}\hat{D}_{_5}\;,
\nonumber\\
&&{\cal W}_{_L}=Y_{_{e_4}}\hat{L}_{_4}\hat{H}_{_d}\hat{E}_{_4}^c+Y_{_{\nu_4}}\hat{L}_{_4}\hat{H}_{_u}\hat{N}_{_4}^c
+Y_{_{e_5}}\hat{L}_{_5}^c\hat{H}_{_u}\hat{E}_{_5}+Y_{_{\nu_5}}\hat{L}_{_5}^c\hat{H}_{_d}\hat{N}_{_5}
\nonumber\\
&&\hspace{1.2cm}
+Y_{_\nu}\hat{L}\hat{H}_{_u}\hat{N}^c+\lambda_{_{N^c}}\hat{N}^c\hat{N}^c\hat{\varphi}_{_L}
+\mu_{_L}\hat{\Phi}_{_L}\hat{\varphi}_{_L}\;,
\nonumber\\
&&{\cal W}_{_X}=\lambda_1\hat{Q}\hat{Q}_{_5}^c\hat{X}+\lambda_2\hat{U}^c\hat{U}_{_5}\hat{X}^\prime
+\lambda_3\hat{D}^c\hat{D}_{_5}\hat{X}^\prime+\mu_{_X}\hat{X}\hat{X}^\prime\;.
\label{superpotential-BL}
\end{eqnarray}
In the superpotential above, the exotic quarks obtain ${\rm TeV}$ scale masses after $\Phi_{_B},\;\varphi_{_B}$
acquiring nonzero VEVs, and the nonzero VEV of $\varphi_{_L}$ implements the seesaw mechanism for the tiny
neutrino masses. Correspondingly, the soft breaking terms are generally given as
\begin{eqnarray}
&&{\cal L}_{_{soft}}={\cal L}_{_{soft}}^{MSSM}-(m_{_{\tilde{N}^c}}^2)_{_{IJ}}\tilde{N}_I^{c*}\tilde{N}_J^c
-m_{_{\tilde{Q}_4}}^2\tilde{Q}_{_4}^\dagger\tilde{Q}_{_4}-m_{_{\tilde{U}_4}}^2\tilde{U}_{_4}^{c*}\tilde{U}_{_4}^c
-m_{_{\tilde{D}_4}}^2\tilde{D}_{_4}^{c*}\tilde{D}_{_4}^c
\nonumber\\
&&\hspace{1.3cm}
-m_{_{\tilde{Q}_5}}^2\tilde{Q}_{_5}^{c\dagger}\tilde{Q}_{_5}^c-m_{_{\tilde{U}_5}}^2\tilde{U}_{_5}^*\tilde{U}_{_5}
-m_{_{\tilde{D}_5}}^2\tilde{D}_{_5}^*\tilde{D}_{_5}-m_{_{\tilde{L}_4}}^2\tilde{L}_{_4}^\dagger\tilde{L}_{_4}
-m_{_{\tilde{\nu}_4}}^2\tilde{\nu}_{_4}^{c*}\tilde{\nu}_{_4}^c-m_{_{\tilde{E}_4}}^2\tilde{e}_{_4}^{c*}\tilde{e}_{_4}^c
\nonumber\\
&&\hspace{1.3cm}
-m_{_{\tilde{L}_5}}^2\tilde{L}_{_5}^{c\dagger}\tilde{L}_{_5}^c
-m_{_{\tilde{\nu}_5}}^2\tilde{\nu}_{_5}^*\tilde{\nu}_{_5}-m_{_{\tilde{E}_5}}^2\tilde{e}_{_5}^*\tilde{e}_{_5}
-m_{_{\Phi_{_B}}}^2\Phi_{_B}^*\Phi_{_B}-m_{_{\varphi_{_B}}}^2\varphi_{_B}^*\varphi_{_B}
\nonumber\\
&&\hspace{1.3cm}
-m_{_{\Phi_{_L}}}^2\Phi_{_L}^*\Phi_{_L}
-m_{_{\varphi_{_L}}}^2\varphi_{_L}^*\varphi_{_L}-\Big(m_{_B}\lambda_{_B}\lambda_{_B}
+m_{_L}\lambda_{_L}\lambda_{_L}+h.c.\Big)
\nonumber\\
&&\hspace{1.3cm}
+\Big\{A_{_{u_4}}Y_{_{u_4}}\tilde{Q}_{_4}H_{_u}\tilde{U}_{_4}^c+A_{_{d_4}}Y_{_{d_4}}\tilde{Q}_{_4}H_{_d}\tilde{D}_{_4}^c
+A_{_{u_5}}Y_{_{u_5}}\tilde{Q}_{_5}^cH_{_d}\tilde{U}_{_5}+A_{_{d_5}}Y_{_{d_5}}\tilde{Q}_{_5}^cH_{_u}\tilde{D}_{_5}
\nonumber\\
&&\hspace{1.3cm}
+A_{_{BQ}}\lambda_{_Q}\tilde{Q}_{_4}\tilde{Q}_{_5}^c\Phi_{_B}+A_{_{BU}}\lambda_{_U}\tilde{U}_{_4}^c\tilde{U}_{_5}\varphi_{_B}
+A_{_{BD}}\lambda_{_D}\tilde{D}_{_4}^c\tilde{D}_{_5}\varphi_{_B}+B_{_B}\mu_{_B}\Phi_{_B}\varphi_{_B}
+h.c.\Big\}
\nonumber\\
&&\hspace{1.3cm}
+\Big\{A_{_{e_4}}Y_{_{e_4}}\tilde{L}_{_4}H_{_d}\tilde{E}_{_4}^c+A_{_{N_4}}Y_{_{N_4}}\tilde{L}_{_4}H_{_u}\tilde{N}_{_4}^c
+A_{_{e_5}}Y_{_{e_5}}\tilde{L}_{_5}^cH_{_u}\tilde{E}_{_5}+A_{_{N_5}}Y_{_{\nu_5}}\tilde{L}_{_5}^cH_{_d}\tilde{N}_{_5}
\nonumber\\
&&\hspace{1.3cm}
+A_{_N}Y_{_N}\tilde{L}H_{_u}\tilde{N}^c+A_{_{N^c}}\lambda_{_{N^c}}\tilde{N}^c\tilde{N}^c\varphi_{_L}
+B_{_L}\mu_{_L}\Phi_{_L}\varphi_{_L}+h.c.\Big\}
\nonumber\\
&&\hspace{1.3cm}
+\Big\{A_1\lambda_1\tilde{Q}\tilde{Q}_{_5}^cX+A_2\lambda_2\tilde{U}^c\tilde{U}_{_5}X^\prime
+A_3\lambda_3\tilde{D}^c\tilde{D}_{_5}X^\prime+B_{_X}\mu_{_X}XX^\prime+h.c.\Big\}\;,
\label{soft-breaking}
\end{eqnarray}
where ${\cal L}_{_{soft}}^{MSSM}$ is the soft breaking terms of MSSM, $\lambda_B,\;\lambda_L$
are gauginos of $U(1)_{_B}$ and $U(1)_{_L}$, respectively.
After the $SU(2)_L$ doublets $H_{_u},\;H_{_d}$ and $SU(2)_L$ singlets $\Phi_{_B},\;\varphi_{_B},\;\Phi_{_L},\;
\varphi_{_L}$ acquiring the nonzero VEVs $\upsilon_{_u},\;\upsilon_{_d},\;\upsilon_{_{B}},\;\overline{\upsilon}_{_{B}}$,
$\upsilon_{_L},\;\overline{\upsilon}_{_L}$
\begin{eqnarray}
&&H_{_u}=\left(\begin{array}{c}H_{_u}^+\\{1\over\sqrt{2}}\Big(\upsilon_{_u}+H_{_u}^0+iP_{_u}^0\Big)\end{array}\right)\;,
\nonumber\\
&&H_{_d}=\left(\begin{array}{c}{1\over\sqrt{2}}\Big(\upsilon_{_d}+H_{_d}^0+iP_{_d}^0\Big)\\H_{_d}^-\end{array}\right)\;,
\nonumber\\
&&\Phi_{_B}={1\over\sqrt{2}}\Big(\upsilon_{_B}+\Phi_{_B}^0+iP_{_B}^0\Big)\;,
\nonumber\\
&&\varphi_{_B}={1\over\sqrt{2}}\Big(\overline{\upsilon}_{_B}+\varphi_{_B}^0+i\overline{P}_{_B}^0\Big)\;,
\nonumber\\
&&\Phi_{_L}={1\over\sqrt{2}}\Big(\upsilon_{_L}+\Phi_{_L}^0+iP_{_L}^0\Big)\;,
\nonumber\\
&&\varphi_{_L}={1\over\sqrt{2}}\Big(\overline{\upsilon}_{_L}+\varphi_{_L}^0+i\overline{P}_{_L}^0\Big)\;,
\label{VEVs}
\end{eqnarray}
the local gauge symmetry $SU(2)_{_L}\otimes U(1)_{_Y}\otimes U(1)_{_B}\otimes U(1)_{_L}$ is broken
down to the electromagnetic symmetry $U(1)_{_e}$.

After the symmetry breaking,  we can obtain the physical spectrum of this model.
The chargino mass matrix is as same as the chargino mass matrix in MSSM.
${Z_ + }, {Z_ - }$ are the matrices to diagonalize the chargino mass mixing matrix ${M_{{{\tilde \chi }^ \pm }}}$
\begin{eqnarray}
Z_ - ^T{\mathcal{M}_{{{\tilde \chi }^ \pm }}}{Z_ + } = diag\Big({m_{{{\tilde \chi_1 }^ \pm }}},{m_{{{\tilde \chi_2 }^ \pm }}}\Big).
\end{eqnarray}
The exotic bottom quark mass matrix is given by
\begin{eqnarray}
\mathcal{M}_{b^{'}}=
\left(\begin{array}{ll}-{1\over\sqrt{2}}\lambda_{_Q}\upsilon_{_B},&-{1\over\sqrt{2}}Y_{_{d_5}}\upsilon_{_u}\\
-{1\over\sqrt{2}}Y_{_{d_4}}\upsilon_{_d},&{1\over\sqrt{2}}\lambda_{_d}\overline{\upsilon}_{_B}
\end{array}\right),
\end{eqnarray}
and this mass matrix is diagonalized by two rotation matrices $W_{_{b^\prime}}$ and $U_{_{b^\prime}}$
\begin{eqnarray}
W_{_{b^\prime}}^\dagger\cdot M_{b^{'}} \cdot U_{_{b^\prime}}={\it diag}\Big(m_{_{b_{1}^\prime}},\;m_{_{b_{2}^\prime}}\Big)\;.
\label{Qmixing-1/3-b}
\end{eqnarray}
The mass matrix of the first three families up-type scalar quark   is given as follow
\begin{eqnarray}
\mathcal{M}_{{{\tilde U}^I}}^2 = \left( {\begin{array}{*{20}{c}}
{m_{\tilde U_L^I}^2 + m_{{U^I}}^2 + D_L^{{{\tilde U}^I}}}+\frac{2}{3}m_{Z_{B}}^{2}\cos2\beta_{B}&{{m_{{U^I}}}({A_{{U^I}}} - {\mu ^ * }\cot \beta )}\\
{{m_{{U^I}}}(A_{{U^I}}^ *  - \mu \cot \beta )}&{m_{\tilde U_R^I}^2 + m_{{U^I}}^2 + D_R^{{{\tilde U}^I}}}-\frac{2}{3}m_{Z_{B}}^{2}\cos2\beta_{B}
\end{array}} \right),
\end{eqnarray}
which has some differences from that of  MSSM,
here  $m_{_{Z_B}}^2=g_{_B}^2(\upsilon_{_B}^2+\overline{\upsilon}_{_B}^2)$ is the mass squared of $U(1)_{B}$ gauge boson $Z_{B}$, and the $D$-terms are
\begin{eqnarray}
&&D_L^{{{\tilde U}^I}} = (\frac{1}{2} - \frac{2}{3}{\sin ^2}{\theta _W})m_Z^2\cos 2\beta,
\nonumber\\
&&D_R^{{{\tilde U}^I}} =  - \frac{2}{3}{\sin ^2}{\theta _W}m_Z^2\cos 2\beta.
\end{eqnarray}
In the basis $(\tilde{Q}_{_4}^{2*},\;\tilde{D}_{_4}^{c},
\;\tilde{Q}_{_5}^{1c},\;\tilde{D}_{_5})$,  the
mass term for the exotic bottom scalar  quarks in the Lagrangian reads as
\begin{eqnarray}
&&-{\cal L}_{_{\tilde{b}^\prime}}^{mass}=
\left(
   \begin{array}{cccc}
     \tilde{Q}_{_4}^{2*}, & \tilde{D}_{_4}^{c}, & \tilde{Q}_{_5}^{1c} &, \tilde{D}_{_5}^{} \\
   \end{array}
 \right)
\cdot {\cal M}_{\tilde{b}^\prime}^2 \cdot \left(
   \begin{array}{cccc}
     \tilde{Q}_{_4}^{2*}, & \tilde{D}_{_4}^{c}, & \tilde{Q}_{_5}^{1c}, & \tilde{D}_{_5}^{} \\
   \end{array}
 \right)^{^{\dagger}}
\label{SQmass-2/3},
\end{eqnarray}
where  ${\cal M}_{\tilde{b}^\prime}^2$ is a $4\times4$ matrix, and  the matrix elements are listed as follows
\begin{eqnarray}
&&({\cal M}_{\tilde{b}^\prime}^2)_{1 1}
=m_{_{\tilde{Q}_4}}^2+{1\over2}Y_{_{u_4}}^2\upsilon_{_u}^2+{1\over2}Y_{_{d_4}}^2\upsilon_{_d}^2
+{1\over2}\lambda_{_Q}^2\upsilon_{_B}^2-\Big({1\over2}-{2\over3}s_{_{\rm W}}^2\Big)m_{_{\rm Z}}^2\cos2\beta
+{B_{_4}\over2}m_{_{Z_B}}^2\cos2\beta_{_B}\;,
\nonumber\\
&&({\cal M}_{\tilde{b}^\prime}^2)_{1 2}=({\cal M}_{\tilde{b}^\prime}^2)^{*}_{2 1}
=-{1\over\sqrt{2}}Y_{_{d_4}}\upsilon_{_d}A_{_{d_4}}+{1\over\sqrt{2}}Y_{_{d_4}}\mu\upsilon_{_d}\;,
\nonumber\\
&&({\cal M}_{\tilde{b}^\prime}^2)_{1 3}=({\cal M}_{\tilde{b}^\prime}^2)^{*}_{3 1}
=-{1\over\sqrt{2}}\lambda_{_Q}\upsilon_{_B}A_{_{BQ}}+\sqrt{2}\lambda_{_Q}\mu_{_B}\overline{\upsilon}_{_B}\;,
\nonumber\\
&&({\cal M}_{\tilde{b}^\prime}^2)_{1 4}=({\cal M}_{\tilde{b}^\prime}^2)^{*}_{4 1}
=-{1\over\sqrt{2}}Y_{_{d_4}}\lambda_d\upsilon_{_d}\overline{\upsilon}_{_B}
+{1\over\sqrt{2}}Y_{_{d_5}}\lambda_Q\upsilon_{_u}\upsilon_{_B}\;,
\nonumber\\
&&({\cal M}_{\tilde{b}^\prime}^2)_{2 2}
=m_{_{\tilde{D}_4}}^2+{1\over2}Y_{_{d_4}}^2\upsilon_{_d}^2
+{1\over2}\lambda_{_d}^2\overline{\upsilon}_{_B}^2-{1\over3}s_{_{\rm W}}^2m_{_{\rm Z}}^2\cos2\beta
-{B_{_4}\over2}m_{_{Z_B}}^2\cos2\beta_{_B}\;,
\nonumber\\
&&({\cal M}_{\tilde{b}^\prime}^2)_{2 3}=({\cal M}_{\tilde{b}^\prime}^2)^{*}_{3 2}
={1\over2}\lambda_Q Y_{_{d_4}}\upsilon_{_d}\upsilon_{_B}+{1\over2}\lambda_dY_{_{d_5}}
\upsilon_{_u}\overline{\upsilon}_{_B}\;,
\nonumber\\
&&({\cal M}_{\tilde{b}^\prime}^2)_{2 4}=({\cal M}_{\tilde{b}^\prime}^2)^{*}_{4 2}
=-{1\over\sqrt{2}}\lambda_{_d}A_{_{BD}}\overline{\upsilon}_{_B}+{1\over\sqrt{2}}\lambda_{_d}\mu_{_B}
\upsilon_{_B}\;,
\nonumber\\
&&({\cal M}_{\tilde{b}^\prime}^2)_{3 3}
=m_{_{\tilde{Q}_5}}^2+{1\over2}Y_{_{u_5}}^2\upsilon_{_d}^2+{1\over2}Y_{_{d_5}}^2\upsilon_{_u}^2
+{1\over2}\lambda_{_Q}^2\upsilon_{_B}^2-\Big({1\over2}+{1\over3}s_{_{\rm W}}^2\Big)m_{_{\rm Z}}^2\cos2\beta
-{1+B_{_4}\over2}m_{_{Z_B}}^2\cos2\beta_{_B}\;,
\nonumber\\
&&({\cal M}_{\tilde{b}^\prime}^2)_{4 4}
=m_{_{\tilde{D}_5}}^2+{1\over2}Y_{_{d_5}}^2\upsilon_{_u}^2
+{1\over2}\lambda_{_d}^2\overline{\upsilon}_{_B}^2+{1\over3}s_{_{\rm W}}^2m_{_{\rm Z}}^2\cos2\beta
+{1+B_{_4}\over2}m_{_{Z_B}}^2\cos2\beta_{_B}\;,
\nonumber\\
&&({\cal M}_{\tilde{b}^\prime}^2)_{3 4}=({\cal M}_{\tilde{b}^\prime}^2)^{*}_{4 3}=
Y_{_{d_5}}A_{_{d_5}}\upsilon_{_u}+{1\over\sqrt{2}}Y_{_{d_5}}\mu\upsilon_{_d}\;.
\label{ESQ-1/3}
\end{eqnarray}
The mass-squared matrix ${\cal M}_{\tilde{b}^\prime}^2$ is  diagonalized  by  the unitary matrix $Z_{\tilde{b}^\prime}$
\begin{eqnarray}
Z_{\tilde{b}^\prime}^{\dag} \cdot {\cal M}_{\tilde{b}^\prime}^2\cdot  Z_{\tilde{b}^\prime}={\it diag}\Big(m_{\tilde{b}_{1}^{\prime}}^{2}, m_{\tilde{b}_{2}^{\prime}}^{2}, m_{\tilde{b}_{3}^{\prime}}^{2}, m_{\tilde{b}_{4}^{\prime}}^{2}\Big),\
\end{eqnarray}
and the physical states are related to the gauge states by
\begin{eqnarray}
\left(
  \begin{array}{c}
    \tilde{b}_{1}^{\prime} \\
    \tilde{b}_{2}^{\prime} \\
    \tilde{b}_{3}^{\prime} \\
    \tilde{b}_{4}^{\prime} \\
  \end{array}
 \right) = Z_{\tilde{b'}}^{\dag}\cdot
 \left(
   \begin{array}{c}
      \tilde{Q}_{_4}^{2} \\
      \tilde{D}_{_4}^{c*} \\
      \tilde{Q}_{_5}^{1c*} \\
      \tilde{D}_{_5}^{*} \\
   \end{array}
 \right).
\end{eqnarray}
The mass squared matrix in the basis $(X^*,~X')$ is
\begin{eqnarray}
\mathcal{M}^{2}_{X}=\left(\begin{array}{cc}\mu_{_X}^2+\frac{1}{2}(\frac{2}{3}+B_{4})m^{2}_{Z_{B}}\cos2\beta_{B} & -\mu_{_X}^*B_{_X}^* \\-\mu_{_X}B_{_X} & \mu_{_X}^2-\frac{1}{2}(\frac{2}{3}+B_{4})m^{2}_{Z_{B}}\cos2\beta_{B} \\ \end{array}\right).
\end{eqnarray}
Adopting the unitary transformation,
the mass eigenstates are
\begin{eqnarray}
~~~\left(     \begin{array}{c}
X_{_1} \\  X_{_2}\\
\end{array}\right) =Z_{_X}^{\dag}\left( \begin{array}{c}
X \\  X'^*\\
\end{array}\right),
\end{eqnarray}
and the mass squared matrix ${M}^{2}_{X}$ is diagonalized by
\begin{equation}
Z^{\dag}_{_X} \cdot M^{2}_{X} \cdot Z_{_X}={\it diag}\Big(m_{_{X1}}^2,\; m_{_{X2}}^2\Big).\
\end{equation}
In four-component Dirac spinors, the mass term for superfields $\tilde{X}$  is given by
\begin{eqnarray}
-{\cal L}_{\tilde{X}}^{mass} = {\mu _X}\tilde{X}\bar{\tilde{X}},
\end{eqnarray}
here, we have defined
\begin{eqnarray}
\tilde{X}=\left(
            \begin{array}{c}
              \psi_{X} \\
              \bar{\psi}_{X^{'}}\\
            \end{array}
          \right)\;.
\end{eqnarray}
So the parameter $\mu_{X}$ is the mass of the particle $\tilde{X}$.

In mass basis, we obtain the couplings of quark-exotic quark and the superfields X
\begin{eqnarray}
\mathcal{L}_{_{Xb'd}}=\sum_{\delta,\epsilon=1}^2\Big(-\lambda_{_1}({W_{b{'}}^\dag})_{_{\delta1}}(Z_{_X})_{_{1\epsilon}}X_{_\epsilon}\bar{b}_{_\delta}P_{_L}d^I
-\lambda_{_3}^*({U_{b{'}}^\dag})_{_{\delta2}}(Z_{_X})_{_{2\epsilon}}X_{_\epsilon}\bar{b}_{_{\delta}}P_{_R}d^I\Big)+h.c.\label{LX}\,.
\end{eqnarray}
We also obtain the couplings of quark-exotic scalar quark and the field $\tilde{X}$
\begin{eqnarray}
\mathcal{L}_{\tilde{X}\tilde b^{'}d^{I}}=\sum_{\rho=1}^{4}\Big({-\lambda _1}(Z_{\tilde{b}^{'}}^{*})_{3\rho} \tilde b_\rho^{'}\tilde{X}{P_L}d^{I}- \lambda _3^*{(Z_{\tilde{b}^{'}})_{4\rho}}\tilde b_\rho^{'}\tilde{X}{P_R}d^{I}\Big)+ h.c.\,,
\end{eqnarray}
where $\lambda _1$, $\lambda _3$ are the coupling coefficients,  and  $\delta,\epsilon,\rho$ are the indices of the flavor.

Considering the radiative corrections, the mass squared matrix for the
neutral CP-even Higgs in the basis $(H_d^0,\;H_u^0)$ is written as \cite{S. P. Li and M. Sher,Y. Okada,J. R. Ellis,R. Barbieri and M. Frigeni,M. Dreesand M. M. Nojiri,M. A. Diaz and H. E. Haber,J. A. Casas,M. S. Carena,S. Heinemeyer1,S. Heinemeyer2,G. Degrassi,M. Frank}
\begin{eqnarray}
&&{\cal M}^2_{even}=\left(\begin{array}{ll}M_{11}^2+\Delta_{11}&M_{12}^2+\Delta_{12}\\
M_{12}^2+\Delta_{12}&M_{22}^2+\Delta_{22}\end{array}\right)\;,
\label{higgs}
\end{eqnarray}
where
\begin{eqnarray}
&&M_{11}^2=m_{_{\rm Z}}^2\cos^2\beta+m_{_{A^0}}^2\sin^2\beta\;,
\nonumber\\
&&M_{12}^2=-(m_{_{\rm Z}}^2+m_{_{A^0}}^2)\sin\beta\cos\beta\;,
\nonumber\\
&&M_{22}^2=m_{_{\rm Z}}^2\sin^2\beta+m_{_{A^0}}^2\cos^2\beta\;,
\nonumber\\
&&\Delta_{11}=\Delta_{11}^{MSSM}+\Delta_{11}^{B}+\Delta_{11}^{L}\;,
\nonumber\\
&&\Delta_{12}=\Delta_{12}^{MSSM}+\Delta_{12}^{B}+\Delta_{12}^{L}\;,
\nonumber\\
&&\Delta_{22}=\Delta_{22}^{MSSM}+\Delta_{22}^{B}+\Delta_{22}^{L}\;,
\nonumber\\
\label{M-CPE1}
\end{eqnarray}
 and the expressions of $\Delta_{11}^{B,L}$, $\Delta_{12}^{B,L}$, $\Delta_{22}^{B,L}$ can be found in Refs. \cite{J. M. Arnold,Tai-Fu Feng}. A Higgs around 125 GeV has been observed at the LHC by ATLAS~\cite{ATLAS} and CMS~\cite{CMS} with the combined significances of 5.9 and 5.0 standard deviations, respectively. So after diagonalizing  the mass squared matrix, the lightest neutral CP even Higgs $m_{_{h_0}}$ should satisfy this constraint. To obtain the Higgs $h_{0}$ with  mass of $125\;{\rm GeV}$ gives a strong  limit on the parameter space. Considering this constraint, we can also obtain $m_{_{A^0}}^2$ from the inverse solution of Eq. (\ref{higgs}). We have
\begin{eqnarray}
&&m_{_{A^0}}^2={m_{_{h_0}}^2(m_{_{\rm z}}^2-m_{_{h_0}}^2+\Delta_{_{11}}+\Delta_{_{22}})-m_{_{\rm z}}^2
\Delta_{_A}+\Delta_{_{12}}^2-\Delta_{_{11}}\Delta_{_{22}}\over -m_{_{h_0}}^2+m_{_{\rm z}}^2\cos^22\beta
+\Delta_{_B}}\;,
\label{Higgs-mass1}
\end{eqnarray}
where
\begin{eqnarray}
&&\Delta_{_A}=\sin^2\beta\Delta_{_{11}}+\cos^2\beta\Delta_{_{22}}+\sin2\beta \Delta_{_{12}}
\;,\nonumber\\
&&\Delta_{_B}=\cos^2\beta\Delta_{_{11}}+\sin^2\beta\Delta_{_{22}}+\sin2\beta \Delta_{_{12}}\;.
\label{Higgs-mass2}
\end{eqnarray}
For the charged Higgs scalars, $H_{1,2}^{\pm}$ are related to the initial Higgs  by the matrix $Z_{H}$,  and the charged Higgs mass $m_{H_{1}^{\pm}}$ satisfy  a  relation with the pseudo-scalar Higgs mass $m_{_{A^0}}$ at tree-level:
\begin{eqnarray}
&&m_{H_{1}^{\pm}}=\sqrt{m_{_{A^0}}^2+m_{_{\rm W}}^{2}}\;,
\label{charged-mass}
\end{eqnarray}
Using the Feynman--t'Hooft gauge, another charged Higgs boson $H_{2}^{\pm}$ has the same mass as  the gauge boson $W$.

\section{${B^0} - {{\bar B}^0}$ mixing\label{sec3}}
When external masses and momenta are neglected, the general form of the effective Hamiltonian for  ${B^0} - {{\bar B}^0}$ mixing at the weak scale can be expressed as\cite{T. Inami and C. S. Lim}
\begin{eqnarray}
{H_{eff}} =\frac{1}{4} \frac{{{\rm{G}}_{\rm{F}}^2}}{{{\pi ^2}}}m_W^2\sum_{\alpha=1}^{8}C_{\alpha}{{{\cal O}_{\alpha}}},
\label{Heff}
\end{eqnarray}
where ${G_F}$ denotes the Fermi constant,  ${C_\alpha } $ are the corresponding Wilson coefficients,
${{{{\cal O}}}_\alpha }$ are the effective operators, which read as
\begin{eqnarray}
&&{\mathcal{O}_1} = \bar d{\gamma _\mu }{P_L}b\bar d{\gamma ^\mu }{P_L}b,
\nonumber\\
&&{\mathcal{O}_2} = \bar d{\gamma _\mu }{P_L}b\bar d{\gamma ^\mu }{P_R}b,
\nonumber\\
&&{\mathcal{O}_3} = \bar d{P_L}b\bar d{P_R}b,
\nonumber\\
&&{\mathcal{O}_4} = \bar d{P_L}b\bar d{P_L}b,
\nonumber\\
&&{\mathcal{O}_5} = \bar d{\sigma _{\mu \nu }}{P_L}b\bar d{\sigma ^{\mu \nu }}{P_L}b,
\nonumber\\
&&{\mathcal{O}_6} = \bar d{\gamma _\mu }{P_R}b\bar d{\gamma ^\mu }{P_R}b,
\nonumber\\
&&{\mathcal{O}_7} = \bar d{P_R}b\bar d{P_R}b,
\nonumber\\
&&{\mathcal{O}_8} = \bar d{\sigma _{\mu \nu }}{P_R}b\bar d{\sigma ^{\mu \nu }}{P_R}b,
\end{eqnarray}
 where ${P_{R,L}} = \left( {1 \pm {\gamma _5}} \right)/2$ denote the chiral projectors, ${\sigma _{\mu \nu }} = \left[ {{\gamma _\mu }, {\gamma _\nu }} \right]/2$,
 the $SU(3)$ color indices here have omitted for simplicity.
\begin{figure}[htbp]
\setlength{\unitlength}{1mm}
\centering
\begin{minipage}[c]{0.5\textwidth}
\includegraphics[width=3.65in]{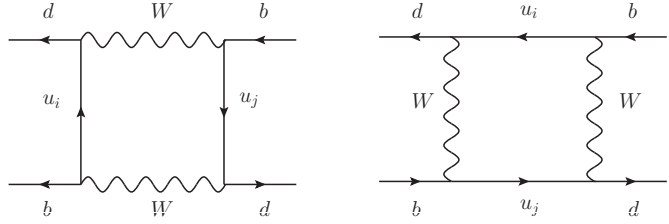}
\end{minipage}%
\caption[]{The box diagram contributions to ${B^0} - {{\bar B}^0}$ mixing in the SM.}
\label{SM}
\end{figure}
\begin{figure}[htbp]
\setlength{\unitlength}{1mm}
\centering
\begin{minipage}[c]{0.5\textwidth}
\includegraphics[width=3.65in]{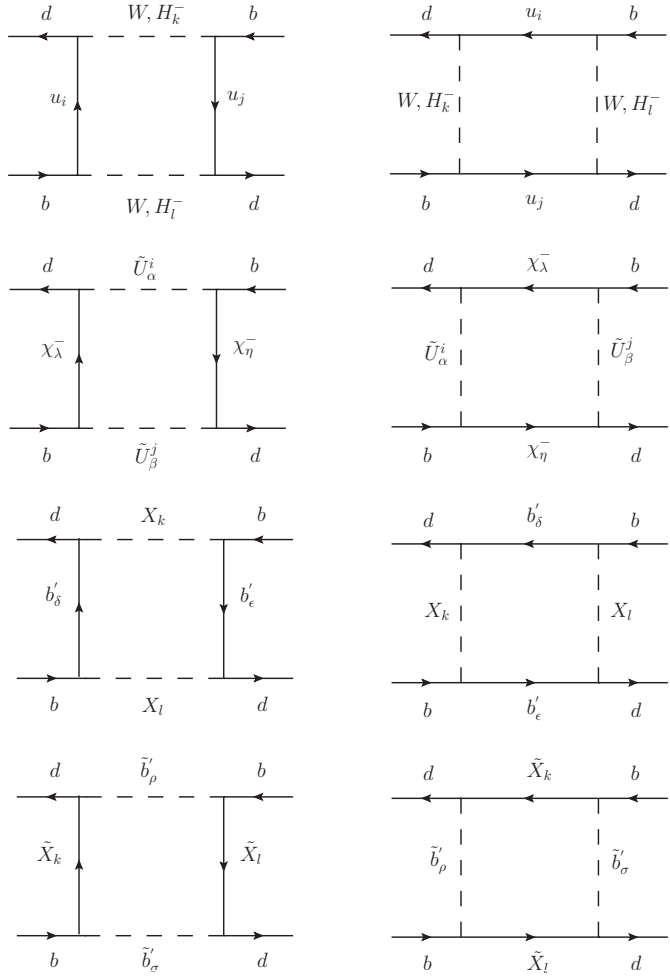}
\end{minipage}
\caption[]{The box diagrams contributing to ${B^0} - {{\bar B}^0}$ mixing in the BLMSSM.}
\label{BLMSSM}
\end{figure}

\newpage

The  box diagram contributions to ${B^0} - {{\bar B}^0}$ mixing  from the SM are displayed in Fig. \ref{SM}, and the box diagrams contributing to ${B^0} - {{\bar B}^0}$ mixing in the BLMSSM are shown in Fig. \ref{BLMSSM}. Note that the diagrams including the particles $\tilde{\chi}$ and $\tilde{X}$ should make a Fierz rearrangement
to ensure that the operators are color singlet  states as follows
\begin{eqnarray}
&&{{\mathcal{O}'}_1} = {\mathcal{O}_1},
\nonumber\\
&&{{\mathcal{O}'}_2} = {\mathcal{O}_2},
\nonumber\\
&&{{\mathcal{O}'}_3} =  - \frac{1}{2}{\mathcal{O}_2},
\nonumber\\
&&{{\mathcal{O}'}_4} =  - \frac{1}{2}{\mathcal{O}_4} - \frac{1}{8}{\mathcal{O}_5},
\nonumber\\
&&{{\mathcal{O}'}_5} = - 6{\mathcal{O}_4} + \frac{1}{2}{\mathcal{O}_5},
\nonumber\\
&&{{\mathcal{O}'}_6} = {\mathcal{O}_6},
\nonumber\\
&&{{\mathcal{O}'}_7} =  - \frac{1}{2}{\mathcal{O}_7} - \frac{1}{8}{\mathcal{O}_8},
\nonumber\\
&&{\mathcal{O}_8}^\prime {\rm{ = }} - 6{\mathcal{O}_7} + \frac{1}{2}{\mathcal{O}_8}.
\end{eqnarray}
The operators with  a   prime  stand for the product of two color non-singlet quark current. After this, the Wilson coefficients are given as follows
\[\begin{array}{l}
{C_1} = {V_{ib}}V_{id}^*{V_{jb}}V_{jd}^*\left( {} \right.{f_{{p^{\rm{2}}}}}({x_{{u_i}}},{x_W},{x_{{u_j}}},{x_W}) - 2\frac{{{x_{{u_i}}}{x_{{u_j}}}}}{{{{\sin }^2}\beta }}{(Z_H^{2k})^2}{f_{\rm{1}}}({x_{{u_i}}},{x_{H_k^ - }},{x_{{u_j}}},{x_W})\\
\\
~~~~~~~ + \frac{1}{4}\frac{{{x_{{u_i}}}{x_{{u_j}}}}}{{{{\sin }^4}\beta }}{(Z_H^{2k})^2}{(Z_H^{2l})^2}{f_{{p^{\rm{2}}}}}({x_{{u_i}}},{x_{H_k^ - }},{x_{{u_j}}},{x_{H_l^ - }}) + Z_{i\alpha }^\lambda Z_{j\beta }^{\lambda *}Z_{j\beta }^\eta Z_{i\alpha }^{\eta *}{f_{{p^{\rm{2}}}}}({x_{\tilde u_\alpha ^i}},{x_{\tilde \chi _\lambda ^ - }},{x_{\tilde u_\beta ^j}},{x_{\tilde \chi _\eta ^ - }})\left. {} \right)\\
\\
~~~~~~~ + \frac{1}{{32G_F^2m_W^4}}|{\lambda _1}{|^4}\left( {} \right.|{(W_{b{'}}^\dag )_{\delta ,1}}{|^2}|{(W_{b{'}}^\dag )_{\varepsilon ,1}}{|^2}|{({Z_X})_{1,l}}{|^2}|{({Z_X})_{1,k}}{|^2}{f_{{p^2}}}({x_{b_\delta ^{'}}},{x_{{X_k}}},{x_{b_\varepsilon ^{'}}},{x_{{X_l}}})\\
\\
~~~~~~~ + |{({Z_{{{\tilde b}^{'}}}})_{3\rho }}{|^2}|{({Z_{{{\tilde b}^{'}}}})_{3\sigma }}{|^2}{f_{{p^2}}}({x_{\tilde X}},{x_{{{\tilde b}^{'}}_\rho }},{x_{\tilde X}},{x_{{{\tilde b}^{'}}_\sigma }})\left. {} \right)
\end{array}\]

\[\begin{array}{l}
{C_{\rm{2}}} = {V_{ib}}V_{id}^*{V_{jb}}V_{jd}^*\left( {} \right.\frac{1}{4}\frac{{\sqrt {{x_b}{x_d}} }}{{{{\sin }^2}\beta {{\cos }^2}\beta }}({x_{{u_i}}} + {x_{{u_j}}})Z_H^{1k}Z_H^{2k}Z_H^{1l}Z_H^{2l}{f_{{p^{\rm{2}}}}}({x_{{u_i}}},{x_{H_k^ - }},{x_{{u_j}}},{x_{H_l^ - }})~~~~~~~~~~~~~~~~~~~~~~~~~~~~~~~~~~~~\\
\\
~~~~~~~ + (Z_{i\alpha }^{d\lambda }Z_{j\beta }^{b\lambda *}Z_{j\beta }^\eta Z_{i\alpha }^{\eta *} + Z_{i\alpha }^\lambda Z_{j\beta }^{\lambda *}Z_{j\beta }^{d\eta }Z_{i\alpha }^{b\eta *}){f_{{p^{\rm{2}}}}}({x_{\tilde u_\alpha ^i}},{x_{\tilde \chi _\lambda ^ - }},{x_{\tilde u_\beta ^j}},{x_{\tilde \chi _\eta ^ - }})\\
\\
~~~~~~~ - 2\sqrt {{x_{\kappa _\lambda ^ - }}{x_{\kappa _\eta ^ - }}} (Z_{i\alpha }^\lambda Z_{j\beta }^{b\lambda *}Z_{j\beta }^{d\eta }Z_{i\alpha }^{\eta *} + Z_{i\alpha }^{d\lambda }Z_{j\beta }^{\lambda *}Z_{j\beta }^\eta Z_{i\alpha }^{b\eta *}){f_{\rm{1}}}({x_{\tilde u_\alpha ^i}},{x_{\tilde \chi _\lambda ^ - }},{x_{\tilde u_\beta ^j}},{x_{\tilde \chi _\eta ^ - }})\left. {} \right)\\
\\
~~~~~~~ + \frac{1}{{64G_F^2m_W^4}}{\left| {{\lambda _1}} \right|^2}{\left| {{\lambda _3}} \right|^2}\left( {} \right.(|{(U_{b{'}}^\dag )_{\delta 2}}{|^2}|{(W_{b{'}}^\dag )_{\varepsilon 1}}{|^2}{({Z_X})_{1,k}}({Z_X})_{2,k}^*({Z_X})_{1,l}^*{({Z_X})_{2,l}}\\
\\
~~~~~~~ + |{(W_{b{'}}^\dag )_{\delta 1}}{|^2}|{(U_{b{'}}^\dag )_{\varepsilon 2}}{|^2}({Z_X})_{1,k}^*{({Z_X})_{2,k}}{({Z_X})_{1,l}}({Z_X})_{2,l}^*){f_{{p^2}}}({x_{b_\delta ^{'}}},{x_{{X_k}}},{x_{b_\varepsilon ^{'}}},{x_{{X_l}}})\\
\\
~~~~~~~ + 2(({Z_{{{\tilde b}^{'}}}})_{3\rho }^ * {({Z_{{{\tilde b}^{'}}}})_{3\sigma }}({Z_{{{\tilde b}^{'}}}})_{4\rho }^*{({Z_{{{\tilde b}^{'}}}})_{4\sigma }} + {({Z_{{{\tilde b}^{'}}}})_{3\rho }}({Z_{{{\tilde b}^{'}}}})_{3\sigma }^ * {({Z_{{{\tilde b}^{'}}}})_{4\rho }}({Z_{{{\tilde b}^{'}}}})_{4\sigma }^*){f_{{p^2}}}({x_{\tilde X}},{x_{{{\tilde b}^{'}}_\rho }},{x_{\tilde X}},{x_{{{\tilde b}^{'}}_\sigma }})\\
\\
~~~~~~~ - 2(|{({Z_{{{\tilde b}^{'}}}})_{3\rho }}{|^2}|{({Z_{{{\tilde b}^{'}}}})_{4\sigma }}{|^2} + |{({Z_{{{\tilde b}^{'}}}})_{3\sigma }}{|^2}|{({Z_{{{\tilde b}^{'}}}})_{4\rho }}{|^2}){x_{\tilde X}}{f_1}({x_{\tilde X}},{x_{{{\tilde b}^{'}}_\rho }},{x_{\tilde X}},{x_{{{\tilde b}^{'}}_\sigma }})\left. {} \right)
\end{array}\]\\

\[\begin{array}{l}
{C_{\rm{3}}} = {V_{ib}}V_{id}^*{V_{jb}}V_{jd}^*\left( {} \right. - 2\frac{{\sqrt {{x_b}{x_d}} }}{{{{\cos }^2}\beta }}{(Z_H^{1k})^2}{f_{{p^{\rm{2}}}}}({x_{{u_i}}},{x_{H_k^ - }},{x_{{u_j}}},{x_W})~~~~~~~~~~~~~~~~~~~~~~~~~~~~~~~~~~~~~~~~~~~~~~~~~~~~~~~~~~~~~~~~~~~\\
\\
~~~~~~~ + \frac{{{x_{{u_i}}}{x_{{u_j}}}\sqrt {{x_b}{x_d}} }}{{{{\sin }^2}\beta {{\cos }^2}\beta }}({(Z_H^{1k})^2}{(Z_H^{2l})^2} + {(Z_H^{2k})^2}{(Z_H^{1l})^2}){f_{\rm{1}}}({x_{{u_i}}},{x_{H_k^ - }},{x_{{u_j}}},{x_{H_l^ - }})\\
\\
~~~~~~~ + 4\sqrt {{x_{\kappa _\lambda ^ - }}{x_{\kappa _\eta ^ - }}} (Z_{i\alpha }^\lambda Z_{j\beta }^{b\lambda *}Z_{j\beta }^{d\eta }Z_{i\alpha }^{\eta *} + Z_{i\alpha }^{d\lambda }Z_{j\beta }^{\lambda *}Z_{j\beta }^\eta Z_{i\alpha }^{b\eta *}){f_{\rm{1}}}({x_{\tilde u_\alpha ^i}},{x_{\tilde \chi _\lambda ^ - }},{x_{\tilde u_\beta ^j}},{x_{\tilde \chi _\eta ^ - }})\left. {} \right)\\
\\
~~~~~~~ + \frac{1}{{16G_F^2m_W^4}}|{\lambda _1}{|^2}|{\lambda _3}{|^2}\left( {} \right.({(W_{b{'}}^\dag )_{\delta 1}}(W_{b{'}}^\dag )_{\varepsilon 1}^*(U_{b{'}}^\dag )_{\delta 2}^*{(U_{b{'}}^\dag )_{\varepsilon 2}}|{({Z_X})_{2,k}}{|^2}|{({Z_X})_{1,l}}{|^2}\\
\\
~~~~~~~ + (W_{b{'}}^\dag )_{\delta 1}^*{(W_{b{'}}^\dag )_{\varepsilon 1}}{(U_{b{'}}^\dag )_{\delta 2}}(U_{b{'}}^\dag )_{\varepsilon 2}^*|{({Z_X})_{1,k}}{|^2}|{({Z_X})_{2,l}}{|^2})\sqrt {{x_{b_\delta ^{'}}}{x_{b_\varepsilon ^{'}}}} {f_1}({x_{b_\delta ^{'}}},{x_{{X_k}}},{x_{b_\varepsilon ^{'}}},{x_{{X_l}}})\\
\\
~~~~~~~ + (|{({Z_{{{\tilde b}^{'}}}})_{3\rho }}{|^2}|{({Z_{{{\tilde b}^{'}}}})_{4\sigma }}{|^2} + |{({Z_{{{\tilde b}^{'}}}})_{3\sigma }}{|^2}|{({Z_{{{\tilde b}^{'}}}})_{4\rho }}{|^2}){x_{\tilde X}}{f_1}({x_{\tilde X}},{x_{{{\tilde b}^{'}}_\rho }},{x_{\tilde X}},{x_{{{\tilde b}^{'}}_\sigma }})\left. {} \right)
\end{array}\]\\

\[\begin{array}{l}
{C_{\rm{4}}} = {V_{ib}}V_{id}^*{V_{jb}}V_{jd}^*\left( {} \right.\frac{{{x_{{u_i}}}{x_{{u_j}}}{x_d}}}{{{{\sin }^2}\beta {{\cos }^2}\beta }}Z_H^{1k}Z_H^{2k}Z_H^{1l}Z_H^{2l}{f_{\rm{1}}}({x_{{u_i}}},{x_{H_k^ - }},{x_{{u_j}}},{x_{H_l^ - }})~~~~~~~~~~~~~~~~~~~~~~~~~~~~~~~~~~~~~~~\\
\\
~~~~~~~ + \sqrt {{x_{\kappa _\lambda ^ - }}{x_{\kappa _\eta ^ - }}} Z_{i\alpha }^{d\lambda }Z_{j\beta }^{\lambda *}Z_{j\beta }^{d\eta }Z_{i\alpha }^{\eta *}{f_{\rm{1}}}({x_{\tilde u_\alpha ^i}},{x_{\tilde \chi _\lambda ^ - }},{x_{\tilde u_\beta ^j}},{x_{\tilde \chi _\eta ^ - }})\left. {} \right)\\
\\
~~~~~~~ + \frac{1}{{32G_F^2m_W^4}}\lambda _1^2\lambda _3^2\left( {} \right.4{(W_{b{'}}^\dag )_{\delta 1}}{(W_{b{'}}^\dag )_{\varepsilon 1}}(U_{b{'}}^\dag )_{\delta 2}^*(U_{b{'}}^\dag )_{\varepsilon 2}^*{({Z_X})_{1,k}}({Z_X})_{2,k}^*{({Z_X})_{1,l}}({Z_X})_{2,l}^*\\
\\
~~~~~~~ \times \sqrt {{x_{b_\delta ^{'}}}{x_{b_\varepsilon ^{'}}}} {f_1}({x_{b_\delta ^{'}}},{x_{{X_k}}},{x_{b_\varepsilon ^{'}}},{x_{{X_l}}}) + ({Z_{{{\tilde b}^{'}}}})_{3\rho }^ * ({Z_{{{\tilde b}^{'}}}})_{3\sigma }^ * ({Z_{{{\tilde b}^{'}}}})_{4\rho }^*({Z_{{{\tilde b}^{'}}}})_{4\sigma }^*{x_{\tilde X}}{f_1}({x_{\tilde X}},{x_{{{\tilde b}^{'}}_\rho }},{x_{\tilde X}},{x_{{{\tilde b}^{'}}_\sigma }})\left. {} \right)
\end{array}\]\\

\[\begin{array}{l}
{C_{\rm{5}}} =  - \frac{1}{4}{V_{ib}}V_{id}^*{V_{jb}}V_{jd}^*\sqrt {{x_{\kappa _\lambda ^ - }}{x_{\kappa _\eta ^ - }}} Z_{i\alpha }^{d\lambda }Z_{j\beta }^{\lambda *}Z_{j\beta }^{d\eta }Z_{i\alpha }^{\eta *}{f_{\rm{1}}}({x_{\tilde u_\alpha ^i}},{x_{\tilde \chi _\lambda ^ - }},{x_{\tilde u_\beta ^j}},{x_{\tilde \chi _\eta ^ - }})~~~~~~~~~~~~~~~~~~~~~~~~~~~~~~~~~~~~~~~~~~~~~~~~~~~~~~~~~~~~~~~~~~~~~~~~~~~~~~~~~~~~~~~~~~~~~~~~~~~~~\\
\\
~~~~~~~ - \frac{1}{{128G_F^2m_W^4}}{\lambda _1}^2{\lambda _3}^2({Z_{{{\tilde b}^{'}}}})_{3\rho }^ * ({Z_{{{\tilde b}^{'}}}})_{3\sigma }^ * ({Z_{{{\tilde b}^{'}}}})_{4\rho }^*({Z_{{{\tilde b}^{'}}}})_{4\sigma }^*{f_1}({x_{\tilde X}},{x_{{{\tilde b}^{'}}_\rho }},{x_{\tilde X}},{x_{{{\tilde b}^{'}}_\sigma }})
\end{array}\]\\

\[\begin{array}{l}
{C_{\rm{6}}} = {V_{ib}}V_{id}^*{V_{jb}}V_{jd}^*\left( {} \right.\frac{1}{4}\frac{{{x_b}{x_d}}}{{{{\sin }^2}\beta {{\cos }^2}\beta }}{(Z_H^{1k})^2}{(Z_H^{1l})^2}{f_{{p^{\rm{2}}}}}({x_{{u_i}}},{x_{H_k^ - }},{x_{{u_j}}},{x_{H_l^ - }})~~~~~~~~~~~~~~~~~~~~~~~~~~~~~~~~~~~~~~~~~~~~~~\\
\\
~~~~~~~ + Z_{i\alpha }^{d\lambda }Z_{j\beta }^{b\lambda *}Z_{j\beta }^{d\eta }Z_{i\alpha }^{b\eta *}{f_{{p^{\rm{2}}}}}({x_{\tilde u_\alpha ^i}},{x_{\tilde \chi _\lambda ^ - }},{x_{\tilde u_\beta ^j}},{x_{\tilde \chi _\eta ^ - }})\left. {} \right)\\
\\
~~~~~~~ + \frac{1}{{32G_F^2m_W^4}}{\left| {{\lambda _3}} \right|^4}\left( {} \right.|{(U_{b{'}}^\dag )_{\delta 2}}{|^2}|{(U_{b{'}}^\dag )_{\varepsilon 2}}{|^2}|{({Z_X})_{2,k}}{|^2}|{({Z_X})_{2,l}}{|^2}{f_{{p^2}}}({x_{b_\delta ^{'}}},{x_{{X_k}}},{x_{b_\varepsilon ^{'}}},{x_{{X_l}}})\\
\\
~~~~~~~ + |{({Z_{{{\tilde b}^{'}}}})_{4\rho }}{|^2}|{({Z_{{{\tilde b}^{'}}}})_{4\sigma }}{|^2}{f_{{p^2}}}({x_{\tilde X}},{x_{{{\tilde b}^{'}}_\rho }},{x_{\tilde X}},{x_{{{\tilde b}^{'}}_\sigma }})\left. {} \right)
\end{array}\]\\

\[\begin{array}{l}
{C_{\rm{7}}} = {V_{ib}}V_{id}^*{V_{jb}}V_{jd}^*\left( {} \right.\frac{{{x_{{u_i}}}{x_{{u_j}}}{x_b}}}{{{{\sin }^2}\beta {{\cos }^2}\beta }}Z_H^{1k}Z_H^{2k}Z_H^{1l}Z_H^{2l}{f_{\rm{1}}}({x_{{u_i}}},{x_{H_k^ - }},{x_{{u_j}}},{x_{H_l^ - }})~~~~~~~~~~~~~~~~~~~~~~~~~~~~~~~~~~~~~\\
\\
~~~~~~~ + \sqrt {{x_{\kappa _\lambda ^ - }}{x_{\kappa _\eta ^ - }}} Z_{i\alpha }^\lambda Z_{j\beta }^{b\lambda *}Z_{j\beta }^\eta Z_{i\alpha }^{b\eta *}{f_{\rm{1}}}({x_{\tilde u_\alpha ^i}},{x_{\tilde \chi _\lambda ^ - }},{x_{\tilde u_\beta ^j}},{x_{\tilde \chi _\eta ^ - }})\left. {} \right)\\
 \\
~~~~~~~ + \frac{1}{{32G_F^2m_W^4}}{\left( {\lambda _1^*} \right)^2}{\left( {\lambda _3^*} \right)^2}\left( {} \right.4(W_{b{'}}^\dag )_{\delta 1}^*(W_{b{'}}^\dag )_{\varepsilon 1}^*{(U_{b{'}}^\dag )_{\delta 2}}{(U_{b{'}}^\dag )_{\varepsilon 2}}({Z_X})_{1,k}^*{({Z_X})_{2,k}}({Z_X})_{1,l}^*{({Z_X})_{2,l}}\\
\\
~~~~~~~ \times \sqrt {{x_{b_\delta ^{'}}}{x_{b_\varepsilon ^{'}}}} {f_1}({x_{b_\delta ^{'}}},{x_{{X_k}}},{x_{b_\varepsilon ^{'}}},{x_{{X_l}}}) + {({Z_{{{\tilde b}^{'}}}})_{3\rho }}{({Z_{{{\tilde b}^{'}}}})_{3\sigma }}{({Z_{{{\tilde b}^{'}}}})_{4\rho }}{({Z_{{{\tilde b}^{'}}}})_{4\sigma }}{f_1}({x_{\tilde X}},{x_{{{\tilde b}^{'}}_\rho }},{x_{\tilde X}},{x_{{{\tilde b}^{'}}_\sigma }})\left. {} \right)
\end{array}\]\\

\[\begin{array}{l}
{C_{\rm{8}}} =  - \frac{1}{4}{V_{ib}}V_{id}^*{V_{jb}}V_{jd}^*\sqrt {{x_{\kappa _\lambda ^ - }}{x_{\kappa _\eta ^ - }}} Z_{i\alpha }^\lambda Z_{j\beta }^{b\lambda *}Z_{j\beta }^\eta Z_{i\alpha }^{b\eta *}{f_{\rm{1}}}({x_{\tilde u_\alpha ^i}},{x_{\tilde \chi _\lambda ^ - }},{x_{\tilde u_\beta ^j}},{x_{\tilde \chi _\eta ^ - }})~~~~~~~~~~~~~~~~~~~~~~~~~~~~~~~~~~~~~~~~~~~~~~~~~~~~~~~~~~~~~~~~~~~~~~~~~~~~~~~~~~~~~~~~~~~~~~~~~~~~~~~~~~~~~~\\
\\
~~~~~~~ - \frac{1}{{128G_F^2m_W^4}}{(\lambda _1^ * )^2}{(\lambda _3^*)^2}{({Z_{{{\tilde b}^{'}}}})_{3i}}{({Z_{{{\tilde b}^{'}}}})_{3j}}{({Z_{{{\tilde b}^{'}}}})_{4i}}{({Z_{{{\tilde b}^{'}}}})_{4j}}{f_1}({x_{\tilde X}},{x_{{{\tilde b}^{'}}_\rho }},{x_{\tilde X}},{x_{{{\tilde b}^{'}}_\sigma }})
\end{array}\]
\begin{eqnarray}
\end{eqnarray}
For convenience,  we have defined the ratio of  mass square as: ${x_i} = m_i^2/m_W^2$, and
here $ Z_{i\alpha }^\lambda $ , $ Z_{i\alpha }^{d\lambda } $... have been
defined as
\begin{eqnarray}
&&Z_{i\alpha }^\lambda {\rm{ = }} - Z_{\tilde U_\alpha ^i}^{1\alpha }Z_ + ^{1\lambda *} + \frac{{\sqrt {2{x_{{u_i}}}} }}{{2\sin \beta }}Z_{\tilde U_\alpha ^i}^{2\alpha }Z_ + ^{2\lambda *}\nonumber\\
&&Z_{i\alpha }^{d\lambda }{\rm{ = }}\frac{{\sqrt {2{x_d}} }}{{2\cos \beta }}Z_{\tilde U_\alpha ^i}^{1\alpha }Z_ - ^{2\lambda }\nonumber\\
&&...
\end{eqnarray}
Here $f_{1}$ and $f_{p^{2}}$ are the functions related to  the one-loop integral functions.
\begin{eqnarray}
\mu^{2\epsilon}\int {\frac{{{d^D}P}}{{{{\left( {2\pi } \right)}^D}}}} \frac{1}{{{p^2} - m_1^2}}\frac{1}{{{p^2} - m_2^2}}\frac{1}{{{p^2} - m_3^2}}\frac{1}{{{p^2} - m_4^2}} = \frac{1}{{16{\pi ^2}m_W^4}}{f_1}\left( {{x_1},{x_2},{x_3},{x_4}} \right)
\end{eqnarray}

\begin{eqnarray}
\mu^{2\epsilon}\int {\frac{{{d^D}P}}{{{{\left( {2\pi } \right)}^D}}}} \frac{1}{{{p^2} - m_1^2}}\frac{1}{{{p^2} - m_2^2}}\frac{1}{{{p^2} - m_3^2}}\frac{1}{{{p^2} - m_4^2}}{p^2} = \frac{1}{{16{\pi ^2}m_W^2}}{f_{{p^2}}}\left( {{x_1},{x_2},{x_3},{x_4}} \right)
\end{eqnarray}
The analytical expressions for the functions ${f_{{p^2}}}\left( {{x_{1}},{x_{2}},{x_{3}},{x_{4}}} \right)$ and ${f_1}\left( {{x_1},{x_2},{x_3},{x_4}} \right)$ are listed in Appendix~\ref{Integral}. It should be noted that we need  perform summation over the repeated indices in the calculations.

The matching scale is chosen  as $\mu _0  = \mu_W$ in our calculations. Now we should evolve the  coefficients from  the scale ${{\mu _W}}$ down to the $B$-meson scale  ${{\mu _b}}$
\begin{eqnarray}
\left[ {\mu \frac{\partial }{{\partial \mu }} + \beta \left( {{\alpha _S}} \right)\frac{\partial }{{\partial {\alpha _S}}} - \frac{{{{\hat \gamma }^T}}}{2}} \right]\vec{C} \left( {\mu ,{\alpha _S}} \right) = 0.
\end{eqnarray}
By solving the remormalization group equation  \cite{M. Ciuchini}, we have
\begin{eqnarray}
{\vec{C}({\mu _{\rm{b}}}) = W({\mu _{\rm{b}}},{\mu _W})\vec{C}({\mu _W})}£¬
\end{eqnarray}
with
\begin{eqnarray}
W({\mu _{\rm{b}}},{\mu _W}) = {\left[ {\frac{{{\alpha _S}({m_W})}}{{{\alpha _S}({m_b})}}} \right]^{\frac{{{\gamma ^{(0)}}}}{{2{\beta _0}}}}},
\end{eqnarray}\\
where $\gamma^{(0)}$ is the anomalous dimensions matrix (ADM) \cite{J.A. Bagger,M. Ciuchini},
and  $\beta_{0}  = \frac{{11{N_c} - 2{n_f}}}{3}$ with ${{N_c}}$ denoting the number of colors and ${{n_f}}$ denoting the
number of active quark flavors.

The mass difference of ${B^0} - {{\bar B}^0}$  mixing can be expressed as
\begin{eqnarray}
\triangle m_{B} = \frac{{\left| {\left\langle {{{\bar B}^0}\left| {{H_{{\rm{eff}}}}\left(
{\Delta B = 2} \right)} \right|{B^0}} \right\rangle } \right|}}{{{m_B}}}.
\end{eqnarray}
After substituting Eq.~(\ref{Heff}) into the above equation, at $B$-meson scale, the mass difference $\triangle m_{B}$ can be written by
\begin{eqnarray}
\triangle m_{B}=\frac{1}{4} \frac{{{\rm{G}}_{\rm{F}}^2}}{{{\pi ^2}}}m_W^2\sum_{\alpha=1}^{8}
\frac{{\left| {C_{\alpha}({\mu _{\rm{b}}})\left\langle {{{\bar B}^0}\left|
{{{{\cal O_{\alpha}}}}({\mu _{\rm{b}}})} \right|{B^0}} \right\rangle } \right|}}{{{m_B}}},
\end{eqnarray}
where, the matrix elements $\left\langle {{{\bar B}^0}\left| {{{\cal
O_{\alpha}}}} \right|{B^0}} \right\rangle$ require non-perturbative QCD calculations  by the  lattice
Monte Carlo estimates. The matrix element is
parameterized as $\left\langle {\bar B^0 \left| {{\cal O}_1 } \right|B^0 } \right\rangle  = \frac{2}{3}{B_B}(\mu )f_B^2 m_B^2$, and the other hadronic matrix elements parameterized are listed in Appendix B.

\section{The numerical analysis \label{sec4}}
In our calculations for the CKM matrix, we apply the Wolfenstein parametrization and set $A = 0.81, ~\lambda  = 0.22, ~\rho  = 0.135, ~\eta  = 0.349$. For the hadronic matrix element, the recent average of the lattice results is ${f_{B_{d}}}\sqrt {{B_{B_{d}}}}  = 216 \pm 15\left( {\rm{MeV}} \right)$ \cite{J. Laiho}, and we adopt the central value of the ${f_{B_{d}}}\sqrt {{B_{B_{d}}}}$ in our calculations. The other SM parameters are chosen as  $m_{W}=80.385~\textrm{GeV}$, $m_{u}=2.3\times10^{-3}~\textrm{GeV}$,
$m_{c}=1.275~\textrm{GeV}$, $m_{t}=173.5~\textrm{GeV}$, $m_{b}=4.18~\textrm{GeV}$, $m_{d}=4.8\times10^{-3}~\textrm{GeV}$, $m_{B}=5.279~\textrm{GeV}$, ${G_F} = 1.166 \times {10^{ - 5}}~{\rm{Ge}}{{\rm{V}}^{ - 2}}$ , ${\alpha _S}({m_W}) = 0.12$ , ${\alpha _S}({m_b}) = 0.22$ \cite{PDG}.

Now we investigate the numerically  behavior of these parameters to the  ${B^0} - {{\bar B}^0}$ mixing in  BLMSSM.  This model contains many parameters.
In our following  discussions,  the  parameters  needed to study  contain $\lambda_{1,3}$, $\mu_{_B}$, $m_{_{Z_B}}$, $m_{D_5}$, $\mu_{_X}$. The other parameters are adopted as Refs. \cite{Tai-Fu Feng,Shu-Min Zhao} which have been  analyzed in the signals of decay channels   $h\rightarrow \gamma\gamma$ and $h\rightarrow VV^{\ast}(V=Z,W)$  with the Higgs mass around 125 GeV.
\begin{eqnarray}
&&m_{_{\tilde{Q}_{1,2,3}}}=m_{_{\tilde{U}_{1,2,3}}}=m_{_{\tilde{D}_{1,2,3}}}=1\;{\rm TeV},\;
\nonumber\\
&&A_{_{d,s,b}}=A_{_{u,c,t}}=-1\;{\rm TeV},\;
\nonumber\\
&&M_{2}=750\;{\rm GeV},\;
\nonumber\\
&&B_4=L_4={3\over2},\;
\nonumber\\
&&\tan\beta=\tan\beta_{B}=\tan\beta_{L}=2,\;
\nonumber\\
&&\mu=-800\;{\rm GeV},\;
\nonumber\\
&&m_{_{\tilde{U}_4}}=m_{_{\tilde{D}_4}}=m_{_{\tilde{U}_5}}
=1\;{\rm TeV},\;\; \nonumber\\
&&m_{_{\tilde{L}_4}}=m_{_{\tilde{\nu}_4}}=m_{_{\tilde{E}_4}}=m_{_{\tilde{L}_5}}=m_{_{\tilde{\nu}_5}}
=m_{_{\tilde{E}_5}}=1\;{\rm TeV}\;,\nonumber\\
&&A_{_{\nu_4}}=A_{_{e_4}}=A_{_{\nu_5}}=A_{_{\nu_5}}=A_{_{u_4}}=A_{_{u_5}}=A_{_{d_4}}=A_{_{d_5}}=550\;{\rm GeV}
\;,\nonumber\\
&&\upsilon_{_{B_t}}=\sqrt{\upsilon_{_B}^2+\overline{\upsilon}_{_B}^2}=3\;{\rm TeV}\;,
\nonumber\\
&&A_{_{BQ}}=A_{_{BU}}=A_{_{BD}}=1\;{\rm TeV},
\nonumber\\
&&Y_{_{u_4}}=0.76Y_t,\;
\nonumber\\
&&Y_{_{d_4}}=Y_{_{u_5}}=0.7Y_b,\;
\nonumber\\
&&Y_{_{d_5}}=0.13Y_t,
\nonumber\\
&&\lambda_{_Q}=\lambda_{_u}=\lambda_{_d}=0.5,\;
\nonumber\\&&m_{_{\nu_4}}=m_{_{\nu_5}}=90\;{\rm GeV},\;\;
\nonumber\\
&&m_{_{e_4}}=m_{_{e_5}}=B_{_X}=100\;{\rm GeV},~\;
\nonumber\\
&&m_{_{\tilde{Q}_4}}=790\;{\rm GeV}.
\label{assumption1}
\end{eqnarray}

\begin{figure}[htbp]
\setlength{\unitlength}{1mm}
\centering
\includegraphics[height=8cm]{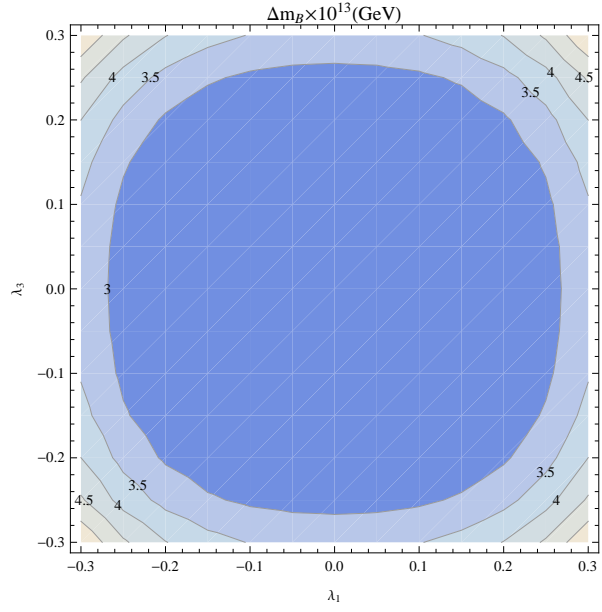}
\caption[]{(Color online) Contour plots of $\Delta {m_B}$  in the parameter space of $\lambda_{1}$ and $\lambda_{3}$.}
\label{lam13}
\end{figure}

In order to see the dependence of the mass difference $\triangle m_{B}$ on the parameters space in the BLMSSM, we fix   $m_{_{\tilde{Q}_5}}=1\;{\rm TeV}$, $m_{_{\tilde{D}_5}}=1\;{\rm TeV}$,   $\mu_{_B}=500~\textrm{GeV}$, $\mu_{_X}=2.4~\textrm{TeV}$, $m_{_{Z_B}}=1\;{\rm TeV}$. From the Wilson coefficients listed in Section \ref{sec3},
one can see that the mass difference ${\triangle}m_{B}$ is the continuous function of the parameters $\lambda_{1}$ and $\lambda_{3}$, and because of the fourth power of $\lambda_{1,3}$ the ${\triangle}m_{B}$  should  remarkably increase with the increasing of  $|\lambda_{1}|$ and $|\lambda_{3}|$ . So $\lambda_{1}$ and $\lambda_{3}$ play an important role to the theoretical prediction on  ${\triangle}m_{B}$. Next, the influence of the parameters $\lambda_{1,3}$  to  ${\triangle}m_{B}$     will  be discussed in detail.
We plot the contours corresponding to the mass difference $\Delta {m_B}$ in the parameter space of $\lambda_{1}$ and $\lambda_{3}$ in Fig. \ref{lam13}. We can see that  $\triangle m_{B}$ increases as $|\lambda_{1,3}|$ increases, and  sensitively depends on $|\lambda_{1,3}|$ when $|\lambda_{1}|$ and $|\lambda_{3}|$ are both larger than 0.2. As one can see,  the values of $|\lambda_{1}|$ and $|\lambda_{3}|$ that  all is larger than 0.25 are disfavored by experiment results under this given assumption.

\begin{figure}[htbp]
\setlength{\unitlength}{1mm}
\centering
\includegraphics[height=8cm]{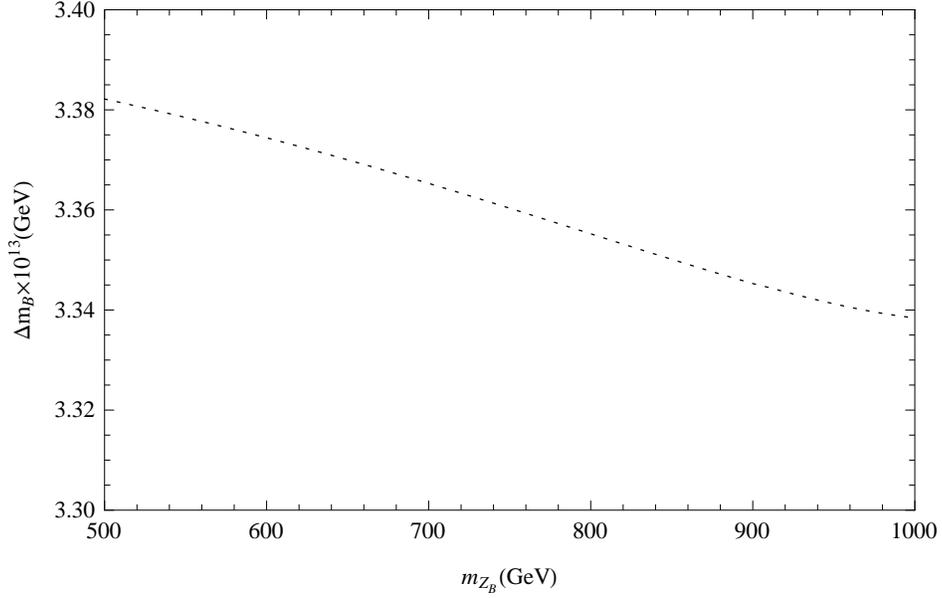}
\caption[]{The mass difference $\Delta {m_B}$ versus the new gauge boson mass ${m_{Z_{B}}}$.}
\label{mZB}
\end{figure}

\begin{figure}[htbp]
\setlength{\unitlength}{1mm}
\centering
\includegraphics[height=8cm]{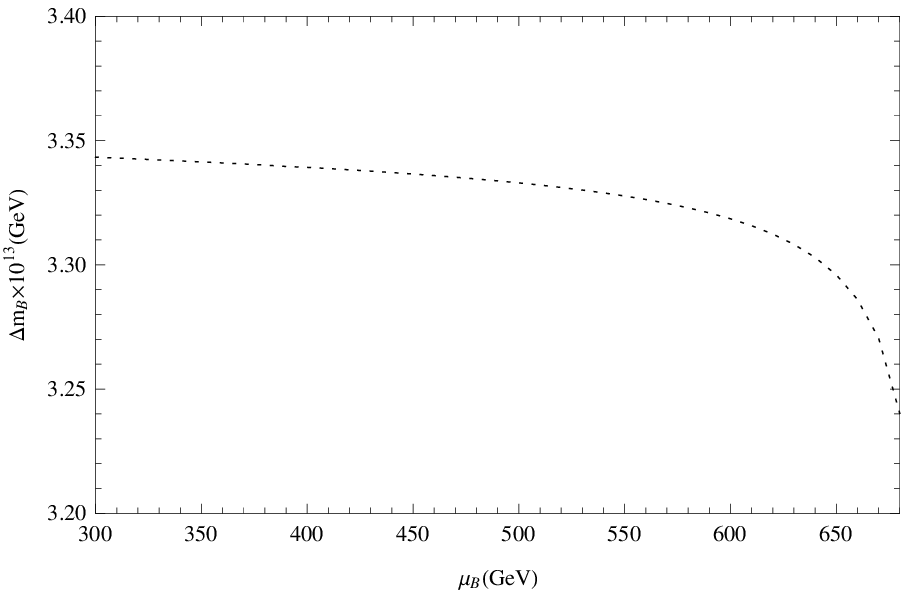}
\caption[]{The mass difference $\Delta {m_B}$ as a function of $\mu_{B}$.}
\label{muB}
\end{figure}

 Next, we investigate the dependence of $\triangle{m_{B}}$ on the  parameter $m_{Z_{B}}$.
 In Fig. \ref{mZB}, we plot  $\triangle{m_{B}}$ varying with the mass of neutral $U(1)_{B}$ gauge boson $Z_{B}$, when $\lambda_{1}=0.25$ and $\lambda_{3}=0.2$. The figure shows that $\triangle{m_{B}}$ decreases as the  $m_{Z_{B}}$ increases. However, it should be noted that the value of the $m_{Z_{B}}$ should not be too large, in order to avoid  some tachyons appearing,  as well as to coincide with the current experimental result on the mass of squarks.   Actually, the corrections of some other parameters to $\triangle{m_{B}}$ are small, such as $m_{\tilde{D_{4}}}$, $m_{\tilde{Q_{4}}}$ and $B_{X}$,  which we would not discuss in this paper.

In the following discussions, we choose $\lambda_{1}=0.2$ for simplicity.
Now, we investigate the dependence of $\triangle{m_{B}}$ on the  parameter $\mu_{B}$.
 Considering that   $\mu_{B}$ is the mass parameter of the "brand new"  Higgs superfields $\Phi_{B}$ and $\phi_{B}$,  the behavior of the  $\triangle{m_{B}}$ versus $\mu_{B}$ when  $\lambda_{3}=0.25$ is shown in Fig. \ref{muB}. The numerical result shows that the  contribution of the parameter $\mu_{B}$  to $\triangle{m_{B}}$ is quite small, when $\mu_{B}$ is lighter than 500 GeV.  When $\mu_{B}$ is heavier than 500 GeV, $\triangle{m_{B}}$ decreases sharply with the increasing of $\mu_{B}$.

\begin{figure}[htbp]
\setlength{\unitlength}{1mm}
\centering
\includegraphics[height=8cm]{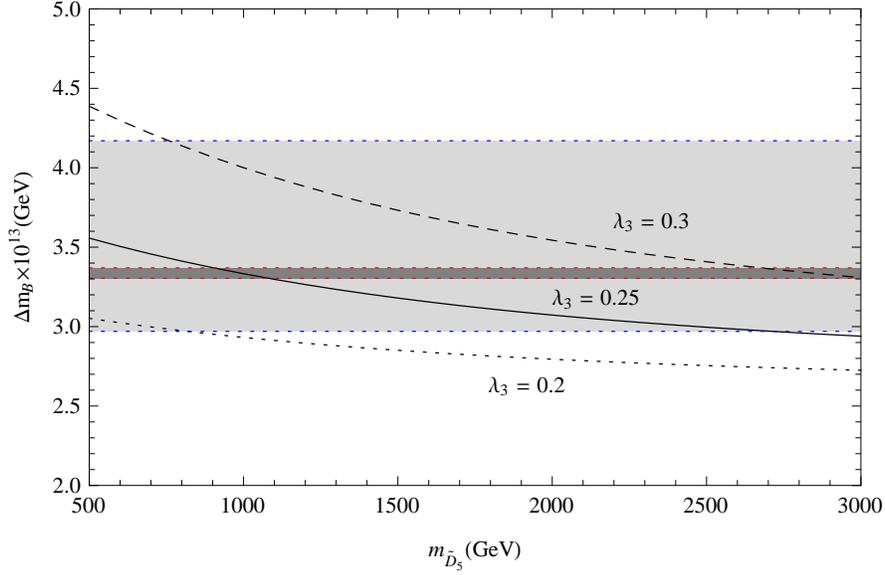}
\caption[]{The mass difference $\Delta {m_B}$ varies with the parameter $m_{\tilde{D_{5}}}$ for three values of $\lambda_{3}$. The light gray area denotes the $\Delta m_B^{SM}$ at 1$\sigma$, and the gray area denotes the $\Delta m_B^{Exp}$ at 1$\sigma$.}
\label{msd5}
\end{figure}

We plot   ${\triangle}m_{B}$ as a
function of the exotic right-handed soft-SUSY-breaking squark mass $m_{\tilde{D_{5}}}$ for three values of $\lambda_{3}$ in Fig. \ref{msd5}, the dotted line
corresponds to the result of $\lambda_{3}=0.2$, the dashed line
corresponds to the result of $\lambda_{3}=0.25$, the dot-dashed line
corresponds to the result of $\lambda_{3}=0.3$.  The light gray area denotes the $\Delta m_B^{SM}$ at 1$\sigma$, and the  gray area denotes the $\Delta m_B^{Exp}$ at 1$\sigma$. As one can see, ${\triangle}m_{B}$  decreases along with the
increasing of $m_{\tilde{D_{5}}}$ for a given value of $\lambda_{3}$.  Fig. \ref{msd5} also exhibits that ${\triangle}m_{B}$  has a strong dependence
on $m_{\tilde{D_{5}}}$ for large values of $\lambda_{3}$.  However, this figure indicates that
the ${\triangle}m_{B}$ declines   slowly with the increasing of $m_{\tilde{D_{5}}}$, when the value  of $\lambda_{3}$ is  small.
Generally speaking,  the influence of the $m_{\tilde{D_{5}}}$ to ${\triangle}m_{B}$ can be neglected as $\lambda_{3}$ is enough small.  Considering the constraint from the $\Delta m_B^{SM}$ at 1$\sigma$, one can see that
small values of $m_{\tilde{D_{5}}}$ can be excluded for large value of $\lambda_{3}$ as well as large values of $m_{\tilde{D_{5}}}$ can be excluded for small value of $\lambda_{3}$ under the given
assumption.

\begin{figure}[htbp]
\setlength{\unitlength}{1mm}
\centering
\includegraphics[height=8cm]{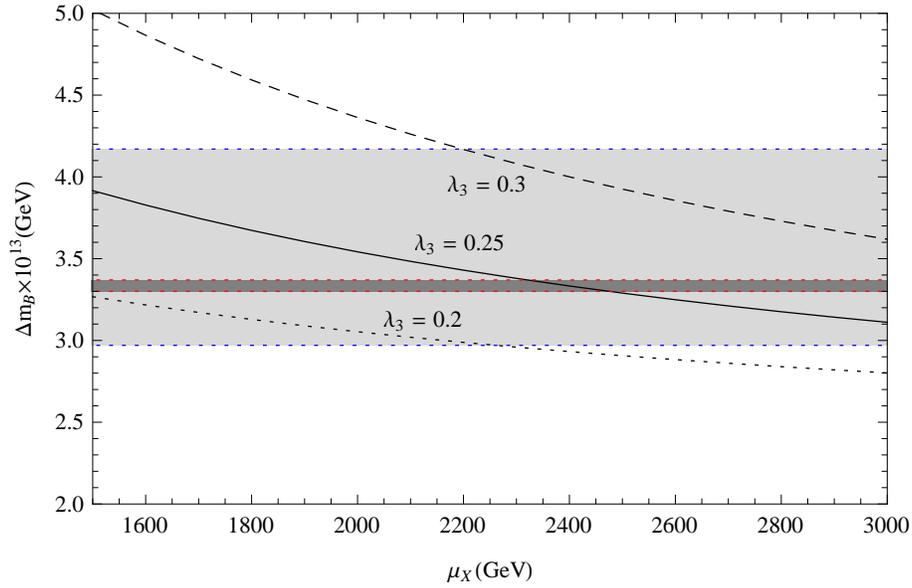}
\caption[]{The mass difference $\Delta {m_B}$ as a function of $\tilde{X}$ mass $\mu_{X}$ for three values of $\lambda_{3}$. The light gray area denotes the $\Delta m_B^{SM}$ at 1$\sigma$, and the  gray area denotes the $\Delta m_B^{Exp}$ at 1$\sigma$.}
\label{mux}
\end{figure}

In Fig. \ref{mux}, we study the dependence  of ${\triangle}m_{B}$ on the particle $\tilde{X}$ mass  $\mu_{X}$. The dotted line corresponds to the result when $\lambda_{3}=0.2$, the solid line corresponds to the result when $\lambda_{3}=0.25$, the dashed line corresponds to the result when $\lambda_{3}=0.3$.   The light gray area denotes the $\Delta m_B^{SM}$ at 1$\sigma$, and the  gray area denotes the $\Delta m_B^{Exp}$ at 1$\sigma$.
It clearly shows a large influence of the new particle $\tilde{X}$ on the mixing of  ${B^0} - {{\bar B}^0}$. The mass difference
${\triangle}m_{B}$  decreases with increasing  of the  $\mu_{X}$    in a very similar manner as that in Fig. \ref{msd5}. We find the mass of the exotic particle $\tilde{X}$ should not be too light for large values of $\lambda_{3}$,
however, the heavy mass of the exotic particle $\mu_{X}$ is also constrained  for small values of $\lambda_{3}$.

\section{CONCLUSIONS\label{sec5}}

With the constraint of a  $125$~GeV  Higgs,
we  analyze  the correction of the extra
fermions and scalars to ${B^0} - {{\bar B}^0}$ mixing in the extension
of the MSSM where baryon number and lepton number are
local gauge symmetries. In this framework, the new particles'
LO correction to ${B^0} - {{\bar B}^0}$ mixing is significant in some parameter space. The numerical evaluations   indicate that
the parameters $\lambda_{1,3}$, $m_{\tilde{D_{5}}}$ and $\mu_{X}$ are sensitive to the process of ${B^0} - {{\bar B}^0}$ mixing. It is well  known  that the space that
is left for hiding some new physics effects in the ${B^0} - {{\bar B}^0}$ mixing is mainly given by the theoretical
error.  With  the development of more precise theoretical analysis (especially the lattice calculations) and accurate experimental measurements,  the  ${B^0} - {{\bar B}^0}$ mixing in the BLMSSM  will   have a  clearer picture and  the parameters space will also be further constrained.

Many experiments have been performed  to search for  baryon number violation (BNV).
Belle and BaBar have  obtained the upper limits on the branching fraction of BNV $\tau$ decays
$\tau^{-}\rightarrow\Lambda \pi^{-}$  and $\tau^{-}\rightarrow\Lambda k^{-}$ \cite{bnv.belle,bnv.babar}.
Some $B$ meson decays $B^{0}\rightarrow\Lambda_{c}^{+}l^{-}$, $B^{-}\rightarrow\Lambda{l^{-}}$ and $B^{-}\rightarrow\bar{\Lambda}{l^{-}}$ have been investigated by BaBar \cite{bnv.babar1}.
Charged lepton flavour violation (CLFV) and BNV decays $\tau^{-}\rightarrow\overline{p}\mu^{+}\mu^{-}$ and $\tau^{-}\rightarrow{p}\mu^{-}\mu^{-}$ have been carried out by LHCb \cite{bnv.lhcb}.
Searching for baryon number violation in top-quark decays has been done by CMS  \cite{bnv.cms}.
However, these experimental searches  for BNV  have yield only upper limits.
On the other hand,  the branching fractions of CLFV process ($\mu\rightarrow{e\gamma}$, $\mu\rightarrow{eee}$, $\tau\rightarrow{l\gamma}$ and $\tau\rightarrow{lll}$ (with $l=e,\mu$), et al.) are predicted very small in the SM. For instance, the SM prediction for  branching fractions in muon decays is smaller than $10^{-50}$. In the BLMSSM, there are some new contributions to these BNV and CLFV processes. And the contributions of BLMSSM may significantly enhance these branching fractions.  One can have BNV signals from the decays of
squarks and gauginos without conflict with   the
current experiments. For instance, if the gluino is the lightest supersymmetric
particle one could have signals with multitops and multibottoms such as $pp\rightarrow \tilde{g}\tilde{g} \rightarrow ttbbjj$ (j stands for a light jet), which may be observed at the LHC \cite{P. F. Perez and M. B. Wise2,P. F. Perez and M. B. Wise11}.
The projected sensitivity for future experiments that searching for the CLFV processes will be largely improved \cite{clfv.B. A. Golden,clfv.J. L. Hewett,clfv.N. Berger,clfv.A. Kurup,clfv.R. J. Abrams,clfv.Y. Kuno,clfv.K. Hayasaka}.  And the running of LHC will resume in 2015 with higher energy and luminosity. So, it would be interesting to investigate this model.
Any observation of BNV or CLFV whose branching fractions is  large than that of  SM prediction would be a clear sign for BSM physics.
Investigating these BNV and CLFV processes can test the BLMSSM and provide constraints on the parameter space.

\section*{Acknowledgements}
The work has been supported by the National Natural Science Foundation of China (NNSFC)
with Grant No. 11275036, No. 11047002, the open project of State
Key Laboratory of Mathematics-Mechanization with Grant No. Y3KF311CJ1, the Natural
Science Foundation of Hebei province with Grant No. A2013201277, and Natural Science Fund of Hebei University with Grant No. 2011JQ05, No. 2012-242.

\appendix

\section{Integral function\label{Integral}}
The functions related to  the one-loop integral functions are given as
\begin{eqnarray}
\begin{array}{l}
{f_1}\left( {{x_1},{x_2},{x_3},{x_4}} \right) = \\
\\
\left\{ \begin{array}{l}
 - \frac{{\log \left( {{x_1}} \right){x_1}}}{{\left( {{x_1} - {x_2}} \right)\left( {{x_1} - {x_3}} \right)\left( {{x_1} - {x_4}} \right)}} + \frac{{\log \left( {{x_2}} \right){x_2}}}{{\left( {{x_1} - {x_2}} \right)\left( {{x_2} - {x_3}} \right)\left( {{x_2} - {x_4}} \right)}}\\
 + \frac{{\log \left( {{x_3}} \right){x_3}}}{{\left( {{x_1} - {x_3}} \right)\left( {{x_3} - {x_2}} \right)\left( {{x_3} - {x_4}} \right)}} + \frac{{\log \left( {{x_4}} \right){x_4}}}{{\left( {{x_1} - {x_4}} \right)\left( {{x_4} - {x_2}} \right)\left( {{x_4} - {x_3}} \right)}},\left( {{x_1} \ne {x_3} ~and~ {x_2} \ne {x_4}} \right)\\
\\
 - \frac{{\log \left( {{x_1}} \right){x_1}}}{{{{\left( {{x_1} - {x_2}} \right)}^2}\left( {{x_1} - {x_3}} \right)}} + \frac{{\log \left( {{x_3}} \right){x_3}}}{{\left( {{x_1} - {x_3}} \right){{\left( {{x_2} - {x_3}} \right)}^2}}}\\
 + \frac{{\log \left( {{x_2}} \right)\left( {x_2^2 - {x_1}{x_3}} \right)}}{{{{\left( {{x_1} - {x_2}} \right)}^2}{{\left( {{x_2} - {x_3}} \right)}^2}}} + \frac{1}{{\left( {{x_1} - {x_2}} \right)\left( {{x_2} - {x_3}} \right)}},\left( {{x_1} \ne {x_3} ~and~ {x_2} = {x_4}} \right)\\
\\
 - \frac{{\log \left( {{x_2}} \right){x_2}}}{{{{\left( {{x_1} - {x_2}} \right)}^2}\left( {{x_2} - {x_4}} \right)}} + \frac{{\log \left( {{x_4}} \right){x_4}}}{{{{\left( {{x_1} - {x_4}} \right)}^2}\left( {{x_2} - {x_4}} \right)}}\\
 + \frac{{\log \left( {{x_1}} \right)\left( {x_1^2 - {x_2}{x_4}} \right)}}{{{{\left( {{x_1} - {x_2}} \right)}^2}{{\left( {{x_1} - {x_4}} \right)}^2}}} - \frac{1}{{\left( {{x_1} - {x_2}} \right)\left( {{x_1} - {x_4}} \right)}},\left( {{x_1} = {x_3} ~and~ {x_2} \ne {x_4}} \right)\\
\\
\frac{{2\log \left( {{x_1}} \right){x_1}{x_2}}}{{{{\left( {{x_1} - {x_2}} \right)}^3}}} - \frac{{2\log \left( {{x_2}} \right){x_1}{x_2}}}{{{{\left( {{x_1} - {x_2}} \right)}^3}}} - \frac{{{x_1} + {x_2}}}{{{{\left( {{x_1} - {x_2}} \right)}^2}}},\left( {{x_1} = {x_3} ~and~ {x_2} = {x_4}} \right)
\end{array} \right.
\end{array}
\end{eqnarray}

\begin{eqnarray}
\begin{array}{l}
{f_{{p^2}}}\left( {{x_1},{x_2},{x_3},{x_4}} \right) = \\
\\
\left\{ \begin{array}{l}
 - \frac{{\log x_1^2}}{{\left( {{x_1} - {x_2}} \right)\left( {{x_1} - {x_3}} \right)\left( {{x_1} - {x_4}} \right)}} + \frac{{\log x_2^2}}{{\left( {{x_1} - {x_2}} \right)\left( {{x_2} - {x_3}} \right)\left( {{x_2} - {x_4}} \right)}}\\
 + \frac{{\log x_3^2}}{{\left( {{x_1} - {x_3}} \right)\left( {{x_3} - {x_2}} \right)\left( {{x_3} - {x_4}} \right)}} + \frac{{\log x_4^2}}{{\left( {{x_1} - {x_4}} \right)\left( {{x_4} - {x_2}} \right)\left( {{x_4} - {x_3}} \right)}},\left( {{x_1} \ne {x_3} ~and~ {x_2} \ne {x_4}} \right)\\
\\
 - \frac{{\log x_1^2}}{{{{\left( {{x_1} - {x_2}} \right)}^2}\left( {{x_1} - {x_3}} \right)}} + \frac{{\log x_3^2}}{{\left( {{x_1} - {x_3}} \right){{\left( {{x_2} - {x_3}} \right)}^2}}}\\
 + \frac{{\log {x_2}\left( {x_2^2 - {x_1}{x_3}} \right)}}{{{{\left( {{x_1} - {x_2}} \right)}^2}{{\left( {{x_2} - {x_3}} \right)}^2}}} + \frac{1}{{\left( {{x_1} - {x_2}} \right)\left( {{x_2} - {x_3}} \right)}},\left( {{x_1} \ne {x_3} ~and~ {x_2} = {x_4}} \right)\\
\\
 - \frac{{\log x_2^2}}{{{{\left( {{x_1} - {x_2}} \right)}^2}\left( {{x_2} - {x_4}} \right)}} + \frac{{\log x_4^2}}{{{{\left( {{x_1} - {x_4}} \right)}^2}\left( {{x_2} - {x_4}} \right)}}\\
 + \frac{{\log {x_1}\left( {x_1^2 - {x_2}{x_4}} \right)}}{{{{\left( {{x_1} - {x_2}} \right)}^2}{{\left( {{x_1} - {x_4}} \right)}^2}}} - \frac{1}{{\left( {{x_1} - {x_2}} \right)\left( {{x_1} - {x_4}} \right)}}{\kern 1pt} {\kern 1pt} ,\left( {{x_1} = {x_3} ~and~ {x_2} \ne {x_4}} \right){\kern 1pt} {\kern 1pt} {\kern 1pt} {\kern 1pt} {\kern 1pt} {\kern 1pt} {\kern 1pt} {\kern 1pt} {\kern 1pt} {\kern 1pt} {\kern 1pt} {\kern 1pt} {\kern 1pt} {\kern 1pt} {\kern 1pt} {\kern 1pt} {\kern 1pt} {\kern 1pt} {\kern 1pt} {\kern 1pt} {\kern 1pt} {\kern 1pt} {\kern 1pt} {\kern 1pt} {\kern 1pt} {\kern 1pt} {\kern 1pt} {\kern 1pt} {\kern 1pt} {\kern 1pt} {\kern 1pt} {\kern 1pt} {\kern 1pt} {\kern 1pt} {\kern 1pt} {\kern 1pt} \\
\\
\frac{{2\log \left( {{x_1}} \right){x_1}{x_2}}}{{{{\left( {{x_1} - {x_2}} \right)}^3}}} - \frac{{2\log \left( {{x_2}} \right){x_1}{x_2}}}{{{{\left( {{x_1} - {x_2}} \right)}^3}}} - \frac{{{x_1} + {x_2}}}{{{{\left( {{x_1} - {x_2}} \right)}^2}}},{\kern 1pt} \left( {{x_1} = {x_3} ~and~ {x_2} = {x_4}} \right)
\end{array} \right.
\end{array}
\end{eqnarray}

\section{Hadronic matrix elements\label{hadronic}}
The hadronic matrix elements can be written as

\[
\begin{array}{l}
 \left\langle {\bar B^0 \left| {{\cal O}_1 } \right|B^0 } \right\rangle  = \frac{2}{3}{B_B}(\mu )f_B^2 m_B^2  \\
  \\
 \left\langle {\bar B^0 \left| {{\cal O}_2 } \right|B^0 } \right\rangle  =  - \frac{1}{{6}}{B_B}(\mu )f_B^2 m_B^2  \\
  \\
 \left\langle {\bar B^0 \left| {{\cal O}_3 } \right|B^0 } \right\rangle  =  - \frac{5}{{12}}{B_B}(\mu )f_B^2 m_B^2  \\
  \\
 \left\langle {\bar B^0 \left| {{\cal O}_4 } \right|B^0 } \right\rangle  = \frac{5}{{12}}{B_B}(\mu )f_B^2 m_B^2  \\
  \\
 \left\langle {\bar B^0 \left| {{\cal O}_5 } \right|B^0 } \right\rangle  = \frac{1}{2}{B_B}(\mu )f_B^2 m_B^2  \\
  \\
 \left\langle {\bar B^0 \left| {{\cal O}_6 } \right|B^0 } \right\rangle  = \frac{2}{3}{B_B}(\mu )f_B^2 m_B^2  \\
  \\
 \left\langle {\bar B^0 \left| {{\cal O}_7 } \right|B^0 } \right\rangle  = \frac{5}{{12}}{B_B}(\mu )f_B^2 m_B^2  \\
  \\
 \left\langle {\bar B^0 \left| {{\cal O}_8 } \right|B^0 } \right\rangle  = \frac{1}{2}{B_B}(\mu )f_B^2 m_B^2.  \\
 \end{array}
\]
Here $f_B$ is the $B$-meson decay constant
constant,  $B_{B}$ is  the  bag parameter.

\end{document}